\documentclass[apj,twocolumn]{aastex63}

\usepackage{graphicx}  % Include figure files
\usepackage{color}     % for preprints
\usepackage{amsmath,amssymb}   % for equations. 
\usepackage{enumitem}
\usepackage{booktabs}
\usepackage{wrapfig}
\usepackage[makeroom]{cancel}
\usepackage[caption=false]{subfig}
\def\BV/{Brunt-V\"{a}is\"{a}l\"{a}}
\shorttitle{Progenitor Explodability}
\shortauthors{Boccioli et al.}

\begin{document}
\doublespace

\title{Explosion mechanism of core-collapse supernovae: role of the Si/Si-O interface}

\correspondingauthor{Luca Boccioli}
\email{lbocciol@nd.edu}

\author[0000-0002-4819-310X]{Luca Boccioli}
\affiliation{Center for Astrophysics, Department of Physics \& Astronomy, University of Notre Dame, 225 Nieuwland Science Hall, Notre Dame, IN 46556, USA}

\author[0000-0003-0390-8770]{Lorenzo Roberti}
\affiliation{Konkoly Observatory, Research Centre for Astronomy and Earth Sciences, E\"otv\"os Lor\'and Research Network (ELKH), Konkoly Thege Mikl\'{o}s \'{u}t 15-17, H-1121 Budapest, Hungary; MTA Centre of Excellence}
\affiliation{Istituto Nazionale di Astrofisica—Osservatorio Astronomico di Roma, Via Frascati 33, I-00040, Monteporzio Catone, Italy}

\author[0000-0002-3164-9131]{Marco Limongi}
\affiliation{Istituto Nazionale di Astrofisica—Osservatorio Astronomico di Roma, Via Frascati 33, I-00040, Monteporzio Catone, Italy}
\affiliation{Kavli Institute for the Physics and Mathematics of the Universe, Todai Institutes for Advanced Study, University of Tokyo, Kashiwa, 277-8583 (Kavli IPMU, WPI), Japan}
\affiliation{INFN. Sezione di Perugia, via A. Pascoli s/n, I-06125 Perugia, Italy}

\author[0000-0002-3164-9131]{Grant J. Mathews}
\affiliation{Department of Physics \& Astronomy, University of Notre Dame, 225 Nieuwland Science Hall, Notre Dame, IN 46556, USA}

\author[0000-0002-3164-9131]{Alessandro Chieffi}
\affiliation{Istituto Nazionale di Astrofisica—Istituto di Astrofisica e Planetologia Spaziali, Via Fosso del Cavaliere 100, I-00133, Roma, Italy}
\affiliation{Monash Centre for Astrophysics (MoCA), School of Mathematical Sciences, Monash University, Victoria 3800, Australia}
\affiliation{INFN. Sezione di Perugia, via A. Pascoli s/n, I-06125 Perugia, Italy}

\begin{abstract}
We present a simple criterion to predict the explodability of massive stars based on the density and entropy profiles before collapse. If a pronounced density jump is present near the Si/Si-O interface, the star will likely explode. We develop a quantitative criterion by using $\sim 1300$ 1D simulations where $\nu$-driven turbulence is included via time-dependent mixing-length theory. This criterion correctly identifies the outcome of the supernova more than $90 \%$ of the time. We also find no difference in how this criterion performs on two different sets of progenitors, evolved using two different stellar evolution codes: FRANEC and KEPLER. The explodability as a function of mass of the two sets of progenitors is very different, showing: (i) that uncertainties in the stellar evolution prescriptions influence the predictions of supernova explosions; (ii) the most important properties of the pre-collapse progenitor that influence the explodability are its density and entropy profiles. We highlight the importance that $\nu$-driven turbulence plays in the explosion by comparing our results to previous works.

\end{abstract}


\section{Introduction}
\label{sec:intro}
The first attempts at describing the physics of core-collapse supernovae (CCSNe) \citep{BBFH_1957,Hoyle1960_SN_Nucleosynthesis} postulated that the explosion could be powered by thermonuclear burning of the material surrounding the core. However, the first numerical simulations  \citep{Colgate_White1966, Arnett1966} identified neutrinos as the primary energy source for the explosion. In these models, the emitted neutrinos traveling outwards will deposit enough energy behind the shock  (the so-called gain region) to energize the explosion started with the core bounce. This is known as the ``prompt neutrino-driven mechanism" since there is no delay between the initial shock expansion and the shock revival caused by neutrino heating.

With a more reliable Equation of State (EOS), neutrino opacities, and numerical algorithms, \cite{Bethe_Wilson1985} found instead that a ``delayed neutrino-heating" after the initial expansion could drive the explosion.  That is, the shock stalls inside the core for a few hundred milliseconds, and is then revived by the neutrinos emitted inside the newly formed proto-neutron star (PNS). However, modern spherically symmetric (1D) codes that employ more accurate EOSs \citep{Lattimer1991_LS, Chabanat1997_SLy4, Shen1998_original_HShen, Hempel2010_HS_RMF, Steiner2013_SFHo, Dutra2014_Skyrme_params, Schneider2017_SROEOS} and neutrino-matter interactions \citep{Bruenn1985, Mezzacappa1993a_infall, Mezzacappa1993b_method, Mezzacappa1993c_nu_e_scat, Thompson2000_mutau_therm, Horowitz2002, BRT2006, Fischer2017_Review_EOS_nu} do not self-consistently explode, except for one zero-metallicity, 9.6 M$_\odot$ progenitor \citep{Melson2015_9.6_expl}.

Finally, with the growth of computing power, the first simulations of CCSNe in two \citep{Miller1993_first2D, Herant1994_first2D} and three \citep{Janka1996_1st_parametric_3D,Fryer2002} spatial dimensions became feasible. Currently, three dimensional simulations routinely explode \citep{Muller2012_2D_GR,Takiwaki2012_original_3DnSNe,Lentz2015_3D,Janka2016_success_expl,Bruenn2016_expl_en,Takiwaki2016_3DnSNe_3D_explosions,OConnor2018_2D_M1,Muller2019_3Dcoconut,Burrows2020_3DFornax,Bugli2021_MHD_3D_SN,Nakamura2022_3DnSNe_binary_star_1987a} and are becoming less computationally demanding. At the same time, axisymmetric (2D) simulations have now a relatively small computational cost compared to three-dimensional (3D) ones. However, the imposed axisymmetry has been shown to artificially enhance turbulence (see for example \cite{Couch2015_turbulence}). 

Therefore, only 3D simulations can ultimately establish what causes an explosion, and they have already shed light on several aspects of the explosion mechanism (cf. review in Ref.~\cite{Muller2016_review}). A clear example that showcases the success of 3D simulations is neutrino-driven turbulent convection, which is a key mechanism in triggering the explosion \citep{Radice2016,Radice2018_turbulence,Mabanta2018_MLT_turb}. Other new phenomena such as the Lepton-number Emission Self-sustained Asymmetry (LESA) \citep{Tamborra2014} and the Standing Accretion Shock Instability (SASI) \citep{Blondin2003} have also been revealed by simulations in three spatial dimensions. However, the impact of these phenomena on the explosion is still a topic of active research.

Nevertheless, 3D simulations currently pose a computational challenge, even for modern supercomputers. Moreover, there is not a good agreement between 3D simulations from different groups \citep{Cabezon2018_3Dcomparison}, except at very small ($<50$ ms) times post bounce. To a lesser extent, this is also true for 2D simulations (see Table 1 from \cite{OConnor2018_2D_M1}), although some promising benchmark work has been done \citep{Pan2019_nu_transport_2D_comparison}. The advantage of 1D simulations is that they run faster and they are consistent across different codes \citep{OConnor2018_comparison}. This makes them an ideal tool to study the vast parameter space of supernovae. The drawback is that the explosion has to be artificially driven.

Different techniques to power an explosion in 1D have been devised over the last few decades \citep{Blinnikov1993_bomb,Woosley1995_pistons,Ugliano2012,Perego2015_PUSH1,Sukhbold2016_explodability,Couch2020_STIR,Ghosh2022_PUSH_EOS}, based upon different parametric models. Moreover, semi-analytical models have been developed over the years \citep{Janka2001_conditions_shk_revival,Muller2016_prog_connection,Summa2016_prog_dependence_vertex,Pejcha2012_antesonic_condition}.  These have attempted to identify what causes one star to explode while another collapses into a black hole. In this paper, we approach the problem of what makes a star explode from an agnostic point of view. We do not assume any pre-determined explosion mechanism, but rather we use the model described in \cite{Boccioli2021_STIR_GR}, whereby the explosion is a natural consequence of $\nu$-driven convection, implemented in 1D using a mixing-length theory (MLT) approach from \cite{Couch2020_STIR}. After determining which stars explode, we ask the question: ``What properties of the progenitor star cause the explosion?". We summarize previous studies addressing this question and inspiring this work in Section \ref{sec:previous_studies}. We describe our numerical setup in Section \ref{sec:num_methods} and the calibration of our models in Section \ref{sec:calibration}. We outline our analysis method in Section \ref{sec:analysis_methods} and present our results in Section \ref{sec:results}, as well as comparisons with previous works. Finally, we present our conclusions in Section \ref{sec:conclusions}.

\section{Previous studies on the explodability problem}
\label{sec:previous_studies}
The question as to what causes an explosion has remained unanswered for many years. Nevertheless, there have been several attempts at describing the explosion process using simple physical arguments. The first efforts are nicely summarized in a review by \cite{Bethe1990_SN_review}. In most of the early work, the focus was on the interior of the PNS, where, if strong convection is present, a large number of neutrinos can be emitted, aiding the shock in its journey towards the outer mantle of the star \citep{Bethe_Wilson1985,WilsonMayle1988,Wilson2005_magnetic_rotating_SN,Wilson2003}. However, it is by now well established that, although convection in the PNS is present, it is not sufficient to liberate the amount of neutrinos necessary to heat the material in the gain region \citep{Bruenn1996_doubly_diffusive,Keil1996_ledoux_PNS,Nagakura2020_PNS_convection}. 

Guided by the numerical results, the first semi-analytical models were also developed. One of these models \citep{Burrows1993_Theory_SN_expl} introduced a key concept that is still in use: the critical luminosity condition. At zeroth order, a supernova can be regarded as being controlled by two parameters: the mass accretion rate $\dot{M}$ and the neutrino luminosity $L_\nu$. For large mass accretion rates more mass falls through the shock, increasing the ram pressure and inhibiting the explosion. At the same time, large mass accretion rates imply more mass being added to the PNS, which increases the neutrino luminosity and therefore neutrino heating, which helps the explosion. 

One can then consider this to be a bifurcation problem. In the $L_\nu-\dot{M}$ plane there is a critical curve dividing solutions that yield explosions (below) and failed supernovae (above). This was further developed in several studies that refined this model by including the effects of $\nu$-driven turbulence and rotation \citep{Janka2001_conditions_shk_revival,Janka2012_review_CCSNe,Muller2016_prog_connection,Pejcha2012_antesonic_condition,Summa2016_prog_dependence_vertex}. It is worth pointing out that the inclusion of $\nu$-driven turbulence was usually done by increasing the shock radius by a factor proportional to the post-shock Mach number. This however fails to capture the complicated dependency of turbulence on the post-bounce dynamics.

Other studies were also performed with the critical condition in mind. Notably, \cite{Ertl2016_explodability} derived a criterion based on the pre-collapse structure of the progenitor.  This was later applied to a wider set of progenitors by \cite{Sukhbold2016_explodability}. Specifically, they defined $M_4$ and $\mu_4$ to be the location in mass and the mass gradient, respectively, of the layer where the entropy crosses a value of $s = 4$ $k_{\rm B}$ baryon$^{-1}$. They then argued that there is a line in the $\mu_4$-$M_4\mu_4$ plane that divides explosions (below) and failed SN (above). This plane is analogous to the $L_\nu-\dot{M}$ plane described above. Using a 1D spherical model to carry out the explosion of several progenitors stars, they indeed found such a separating curve. More details of their model in comparison to ours are given in Section \ref{sec:ertl_comparison}.

Another attempt at understanding the behavior of supernova explosions was done by \cite{OConnor2011_explodability}, who introduced the compactness parameter $\xi_{\rm M}$
\begin{equation}
    \label{eq:comp_25}
    \xi_{\rm M} = \left. \frac{M/{\rm M}_\odot}{R(M_{\rm bary}=M) / 1000 {\rm km}} \right|_{t = t_{\rm bounce}} ~,
\end{equation}
where M is the value of the mass at which this parameter should be evaluated. They used fully general relativistic simulations to study in detail the trajectory of the shock until black hole formation.  They then derived a criterion that connects $\xi_{2.5}$ to the time it takes to form a black hole. 

Large $\xi_{2.5}$ leads to rapid black hole formation since the mass of the PNS increases very quickly due to the large mass accretion rates. Then, by artificially enhancing the neutrino luminosity (a very crude but easy way to achieve an explosion in 1D) they derived a criterion stating that stars with $\xi_{2.5} < 0.45$ explode, and stars with $\xi_{2.5} > 0.45$ collapse to black holes. 

This work was later expanded to study the dependency of black hole formation on the EOS \citep{Schneider2020_EOS_dependence_BH}. Since 1D models are well suited to a study of the collapse phase, the results regarding black hole formation are very reliable. However, $\xi_{2.5}$ does not contain enough details to properly account for all of the physical mechanisms in effect during the explosion. Therefore, it cannot accurately predict the outcome of the supernova.

A new tool that models $\nu$-driven convection was recently developed by \cite{Couch2020_STIR}, using a time-dependent mixing-length theory approach that can be implemented in 1D simulations. Therefore, neutrino heating is increased by a physically motivated mechanism, seen in 3D simulations, rather than by an artificial increase of the neutrino luminosity. Using this model one can, within the uncertainties of mixing-length theory, recover approximately the same shock dynamics seen in 2D and 3D simulations, but at a much smaller computational cost. This inspired us to perform simulations for a wide range of progenitors and then, guided by several of the studies mentioned above, find the dynamical properties of such simulations that could explain the outcome of the explosion. Finally, by connecting these dynamical properties to the thermodynamic structure of the pre-collapse progenitor, one can formulate a criterion that predicts the outcome of the supernova without the need of performing the simulations.

\section{Numerical Methods}
\label{sec:num_methods}
For this work, we used two sets of stellar models for massive stars: one set of progenitors based upon the KEPLER code \citep{Sukhbold2016_explodability}, and one set based upon the FRANEC code \citep{Chieffi2020_presupernova_models}. For all the figures in this paper, we identify FRANEC progenitors with circles and KEPLER progenitors with squares. We highlight that the pre-supernova progenitors from KEPLER suffer from a known bug that alters neutrino cooling. The bug was later fixed in \cite{Sukhbold2018_preSN_KEPLER_bug} and was shown to have a small impact on the final pre-supernova structure. Despite this bug, we still decided to use the old pre-supernova progenitors, so that we could compare our results to the works of \cite{Sukhbold2016_explodability} and \cite{Ertl2016_explodability}.

To model the collapse and subsequent explosion of the progenitor stars, we used the open-source code \texttt{GR1D} \citep{OConnor2010,OConnor2015}. It is a spherically-symmetric, fully generally relativistic hydrodynamic code coupled to a neutrino radiation transport module that solves the Boltzmann equation using a two-moment (M1) scheme. We included the effects of neutrino-driven turbulent convection through a relativistic version of the time-dependent MLT model of \textit{STIR} \cite{Couch2020_STIR}, as described in \cite{Boccioli2021_STIR_GR}. All of the simulations were performed using 700 radial zones extending out to $15000$ km, with a uniform spacing of 0.3 km for the inner 20 km and a logarithmic spacing outside. For the neutrino opacities, we used the tables from NuLib \citep{OConnor2015}, with 18 neutrino energy groups from 1 MeV to $\sim 280$ MeV, and the same neutrino-matter interactions used in \cite{Boccioli2022_EOS_effect}. All of the simulations were performed using the SFHo equation of state \citep{Steiner2013_SFHo}.

\section{Calibration of \textit{STIR}}
\label{sec:calibration}
\begin{figure*}
    \centering
    \includegraphics[width=\textwidth]{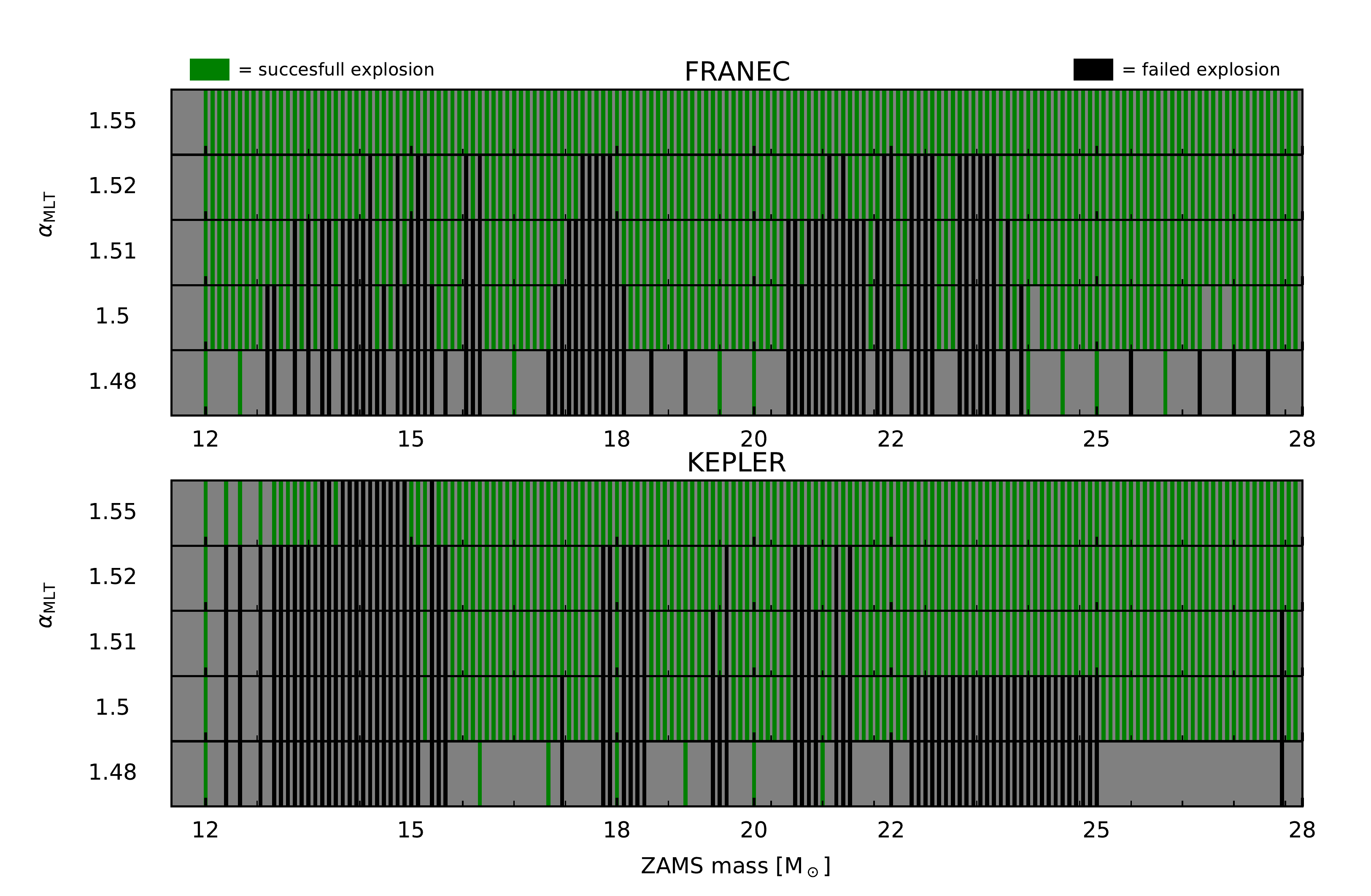}
    \caption{Explodability as a function of progenitor mass for different values of $\alpha_{\rm MLT}$ for the FRANEC progenitors (top) and KEPLER progenitors (bottom). Simulations resulting in successful shock revival are shown in green, and simulations resulting in a failed SN are shown in black. Not every band corresponds to an actual simulation. In some cases, the outcome of the simulation (i.e. the color of the band) is inferred from simulations of the same progenitors at different values of $\alpha_{\rm MLT}$. If a star explodes for a given value of $\alpha_{\rm MLT}$ it will also explode for larger values. Conversely, if a star results in a failed SN for a given value of $\alpha_{\rm MLT}$ it will also fail to explode for smaller values. Grey regions correspond to stars that were not simulated and no information could be inferred from results at different values of $\alpha_{\rm MLT}$ }
    \label{fig:calibration_explodability}
\end{figure*}

\begin{figure}
    \centering
    \includegraphics[width=\linewidth]{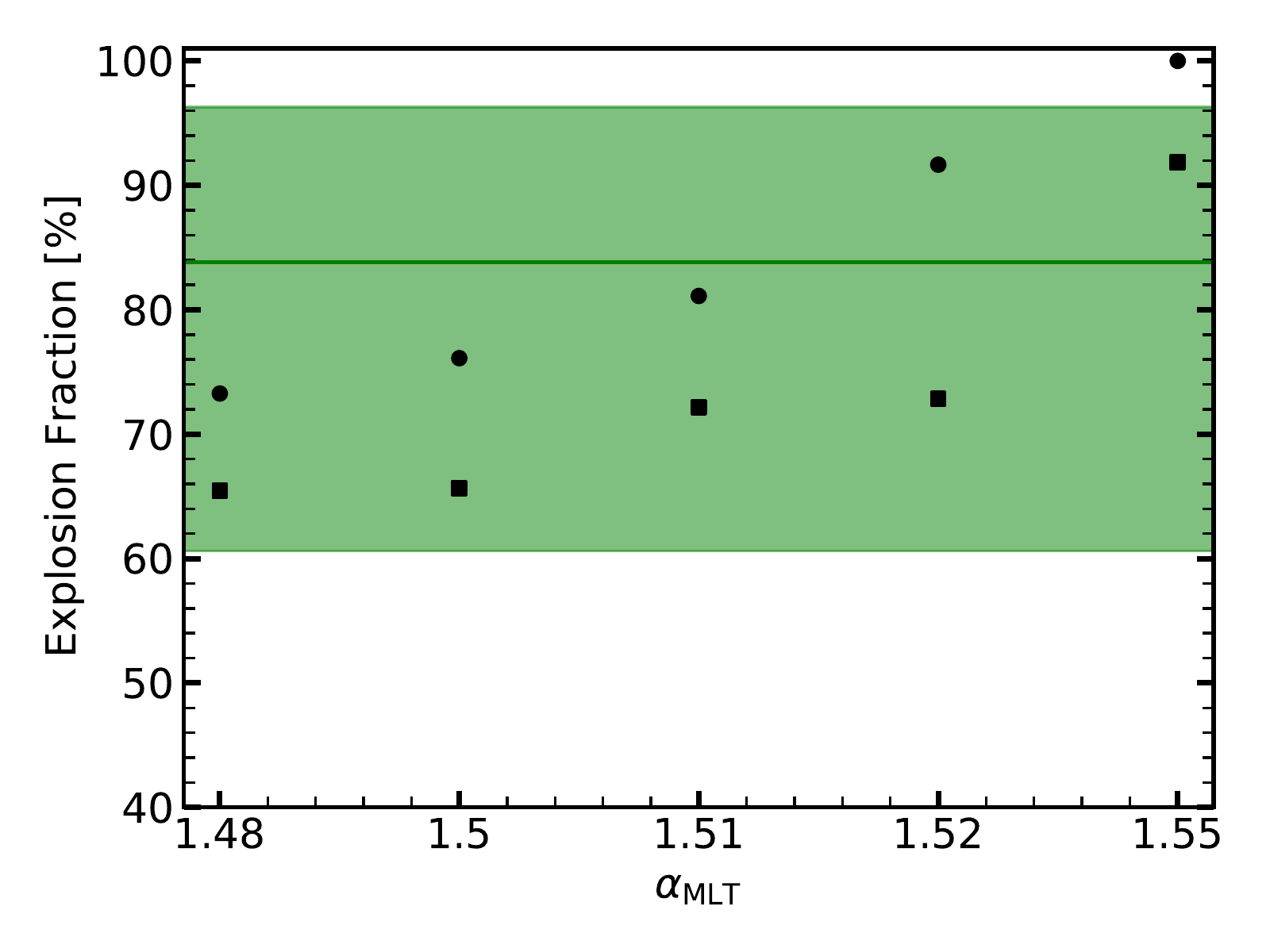}
    \caption{The explosion fraction is calculated by weighing the explodability shown in Figure \ref{fig:calibration_explodability} with a Salpeter initial mass function. All stars from 9 to 12 M$_\odot$ are assumed to result in explosions, consistent with results from multi-dimensional simulations. The black symbols are the explosion fractions assuming the explodability shown in Figure \ref{fig:calibration_explodability} for FRANEC (circles) and KEPLER (squares) progenitors, for different values of $\alpha_{\rm MLT}$. The latest observational data from \cite{Neustadt2021_failedSN_frac} estimate the fraction of failed supernovae to be $f_{\rm fSNe} = 0.16^{+0.23}_{-0.12}$ with a 90\% confidence. The interval and median values are represented as a shaded green region and horizontal line. Notice that the explosion fraction is $1 - f_{\rm fSNe}$}
    \label{fig:explosion_fraction}
\end{figure}

As mentioned in the previous section, in our 1D simulations the explosion can be achieved by virtue of \textit{STIR}, a time-dependent MLT model for $\nu$-driven turbulent convection. Therefore, the amount of turbulence generated (and subsequently dissipated) depends on a parameter $\alpha_{\rm MLT}$ of order unity. For more details about this model, we refer the reader to \cite{Mabanta2018_MLT_turb}, \cite{Couch2020_STIR} and \cite{Boccioli2021_STIR_GR}.

As shown in \cite{Boccioli2021_STIR_GR}, the spatial and neutrino energy resolutions change the value of $\alpha_{\rm MLT}$ that best reproduces the results of 3D simulations. Therefore, we chose to adopt the same numerical setup (described in \ref{sec:num_methods}) as in \cite{Boccioli2021_STIR_GR} and \cite{Boccioli2022_EOS_effect}. There, a value of $\alpha_{\rm MLT}$ around 1.5 was selected, based upon comparisons with 3D simulations and observational constraints on the fraction of stars that collapse to black holes. For this paper, we identify the best-fit range $1.5 \leq \alpha_{\rm MLT} \leq 1.52$. 

To justify the range chosen, we show in Figure \ref{fig:calibration_explodability} the explodability as a function of zero-age main sequence (ZAMS) mass for different values of $\alpha_{\rm MLT}$, extending slightly above and below our best-fit range. Then, we show in Figure \ref{fig:explosion_fraction} the corresponding explosion fractions. Only simulations with $\alpha_{\rm MLT} \geq 1.5$ lie within the observational constraint from \cite{Neustadt2021_failedSN_frac}, depicted as a green-shaded region. Since we don't have simulations of progenitors with masses $< 12$ M$_\odot$, to estimate the explosion fractions we assume that every progenitor with a mass lower than 12 M$_\odot$ explodes. This assumption is justified by many 3D simulations of low-mass progenitors, as well as by the 1D explodability studies by \cite{Sukhbold2016_explodability}, \cite{Couch2020_STIR}, and \cite{Boccioli2022_EOS_effect}. 

Additionally, values that are somewhat larger than $\alpha_{\rm MLT} = 1.5$ are not compatible with the comparison to 3D simulations performed in \cite{Boccioli2021_STIR_GR} and \cite{Boccioli2022_EOS_effect}. Specifically, the results with $\alpha_{\rm MLT} = 1.55$ in \cite{Boccioli2022_EOS_effect} deviate substantially from the 3D data. Also, Figure 2 in \cite{Boccioli2021_STIR_GR} shows that $\alpha_{\rm MLT} = 1.5$ yields the best match to the 3D explosion properties. However, those simulations were run with a higher spatial resolution, more neutrino energy groups, and slightly different neutrino opacities. Therefore, as discussed in the appendix of \cite{Boccioli2021_STIR_GR}, it is expected that a larger $\alpha_{\rm MLT}$ would be required for the simulations in this paper. Thus, we chose our upper limit to be $\alpha_{\rm MLT} = 1.52$. 

Further justification is provided by the fact that, for FRANEC progenitors, $\alpha_{\rm MLT} = 1.52$ is already near the upper limit for explosion fraction, as can be seen in Figure \ref{fig:explosion_fraction}. On the other hand, one could make the argument that the calibration for $\alpha_{\rm MLT}$ can be different for KEPLER and FRANEC progenitors. However, $\nu$-driven convection only depends on the mass accretion rate in the gain region and the neutrino luminosity emitted from the PNS. Therefore, there is no reason to assume that differences in the stellar evolution codes would lead to different $\nu$-driven convection. This is confirmed by our simulations, where values around $\alpha_{\rm MLT} \approx 1.5$ are compatible with the observed fraction of failed SNe for both progenitor sets. To summarize, we identify our best-fit range to be $1.5 \leq \alpha_{\rm MLT} \leq 1.52$, and our best-fit value to be $\alpha_{\rm MLT} = 1.51$.

It's important to point out that Figure \ref{fig:calibration_explodability} does not only show the outcome of the supernova for progenitors that were simulated but also the outcome inferred by simulations of that progenitors at different values of $\alpha_{\rm MLT}$. In other words, if a star explodes for a given value of $\alpha_{\rm MLT}$, it will also explode for larger values. Similarly, if a star does not explode for a given value of $\alpha_{\rm MLT}$, it will also not explode for smaller values. We can therefore avoid running many simulations since, for this study, we are mainly interested in the outcome of the supernova rather than in the details of the post-bounce dynamics.

As can be seen from Figure \ref{fig:explosion_fraction}, the explodability is a steep function of $\alpha_{\rm MLT}$. This means that several progenitors do not explode at a given value of $\alpha_{\rm MLT}$ but do explode for a slightly larger value. To quantify how ``close" to an explosion a progenitor that results in a failed SN is, we consider the advection and heating timescales. 

The advection timescale $\tau_{\rm adv}$ is a measure of how much time the infalling material spends in the gain region before settling onto the PNS. The heating timescale $\tau_{\rm heat}$ indicates how long it takes for neutrinos to deposit energy in the gain region. It is expected \citep{Janka2001_conditions_shk_revival,Buras2006_2D_diag_ene,Radice2018_turbulence} that for a ratio $\tau_{\rm adv} / \tau_{\rm heat} \gtrsim 1$ the explosion becomes favorable since the matter in the gain region is exposed to neutrino heating for a long time before it can settle onto the PNS. 

This is conceptually the same as defining a critical luminosity condition, since larger $\dot{M}$ will correspond to smaller $\tau_{\rm adv}$, and larger $L_\nu$ will correspond to smaller $\tau_{\rm heat}$. In our 1D simulations, we find that some failed SN have $\tau_{\rm adv} / \tau_{\rm heat} > 1$. Therefore, we expect those stars to be on the verge of shock revival, and a slight increase of $\alpha_{\rm MLT}$ would result in explosions.

There are several equivalent definitions of these timescales \citep{Buras2006_2D_diag_ene,Marek2009_SASI_diag_ene,Muller2012_2D_GR,Fernandez2012_timescales,Radice2018_turbulence}, and particular care has to be applied in the definition of $\tau_{\rm heat}$. We define the advection timescale to be:
\begin{equation}
    \tau_{\rm adv} = \frac{M_{\rm gain}}{\dot{M}}~,
\end{equation}
where $M_{\rm gain}$ is the total mass in the gain region and $\dot{M}$ is the mass accretion rate calculated at 500 km. 

The heating timescale is taken to be:
\begin{equation}
    \tau_{\rm heat} = \frac{\lvert E_{\rm gain} \rvert}{\dot{\mathcal{Q}}_{\rm net}}~,
\end{equation}
where $\dot{\mathcal{Q}}_{\rm net}$ is the net neutrino heating rate in the gain region and $E_{\rm gain}$ is the binding energy in the gain region, defined as in \cite{Muller2012_2D_GR}:

\begin{equation}
    \label{eq:Egain}
    E_{\rm gain} = \int_{\rm gain} \left\{\alpha\left[ ( \rho + \rho\epsilon_{\rm th} + P) W^2 - P\right] - \rho W \right\} {\rm d}\widetilde{V}~.
\end{equation}
Here $\alpha$ is the lapse function, $\rho$ is the density, $\epsilon_{\rm th}$ is the specific internal thermal energy, $P$ is the pressure, $W$ is the Lorentz factor and $\widetilde{V}$ is the proper volume. Notice that Equation \eqref{eq:Egain} does not include the recombination energy of the matter, which might be a significant contribution to the explosion energy \citep{Bruenn2016_expl_en}. However, to calculate the heating timescale, one only needs the total energy of the matter at that specific time and thermodynamic conditions, and therefore the recombination energy should not be included \citep{Fernandez2012_timescales,Radice2016}.

In the evaluation of the timescales, it is very important to use the correct expression for $\epsilon_{\rm th}$. Usually, the internal energy provided by the EOS includes the binding energy of nuclear matter and therefore differs from the actual thermal energy. An approximate way to calculate $\epsilon_{\rm th}$ is to calculate the internal energy from the EOS $\epsilon(T,\rho,Y_{\rm e})$ minus the internal energy for the same density and electron fraction but at zero temperature $\epsilon(0,\rho,Y_{\rm e})$. 

This definition of $\epsilon_{\rm th} = \epsilon(T,\rho,Y_{\rm e}) - \epsilon(0,\rho,Y_{\rm e})$ is accurate enough if one needs a rough estimate of the diagnostic explosion energy \citep{Muller2012_2D_GR,Betranhandy2020_pair,Boccioli2021_STIR_GR}, but should not be applied to the calculation of the heating timescale. For that, we resort to the definition of $\epsilon_{\rm th}$ used by \cite{Bruenn2016_expl_en} and \cite{Harada2020_2D_SN_EoS_effect}:
\begin{equation}
    \label{eq:eps_th}
    \epsilon_{\rm th} \equiv \rho \frac{3}{2}\frac{k_{\rm B} T}{m_{\rm u}} \frac{X_{\rm h}}{A_{\rm h}} + aT^4 + \left( \epsilon_{e^-e^+} - \rho Y_{\rm e} \frac{m_{\rm e}c^2}{m_{\rm u}} \right).
\end{equation}

The first term in Equation \ref{eq:eps_th} is the contribution of the heavy nuclear species treated as an ideal gas, where $X_{\rm h}$ and $A_{\rm h}$ are the mass fraction and number of nucleons of a representative heavy nucleus, as provided by the EOS table. The second term is the contribution from an ideal photon gas, where $a$ is the radiation constant and $T$ is the temperature. The third term is the internal energy of the electron-positron gas minus the contribution from the rest mass of the electron. Here, $m_{\rm u}$ is the atomic mass unit, and $k_{\rm B}$ is the Boltzmann constant.

\section{Shock dynamics during the post-bounce phase}
\label{sec:analysis_methods}
\subsection{Accretion of a sharp density gradient}
\begin{figure*}
    \centering
    \gridline{\fig{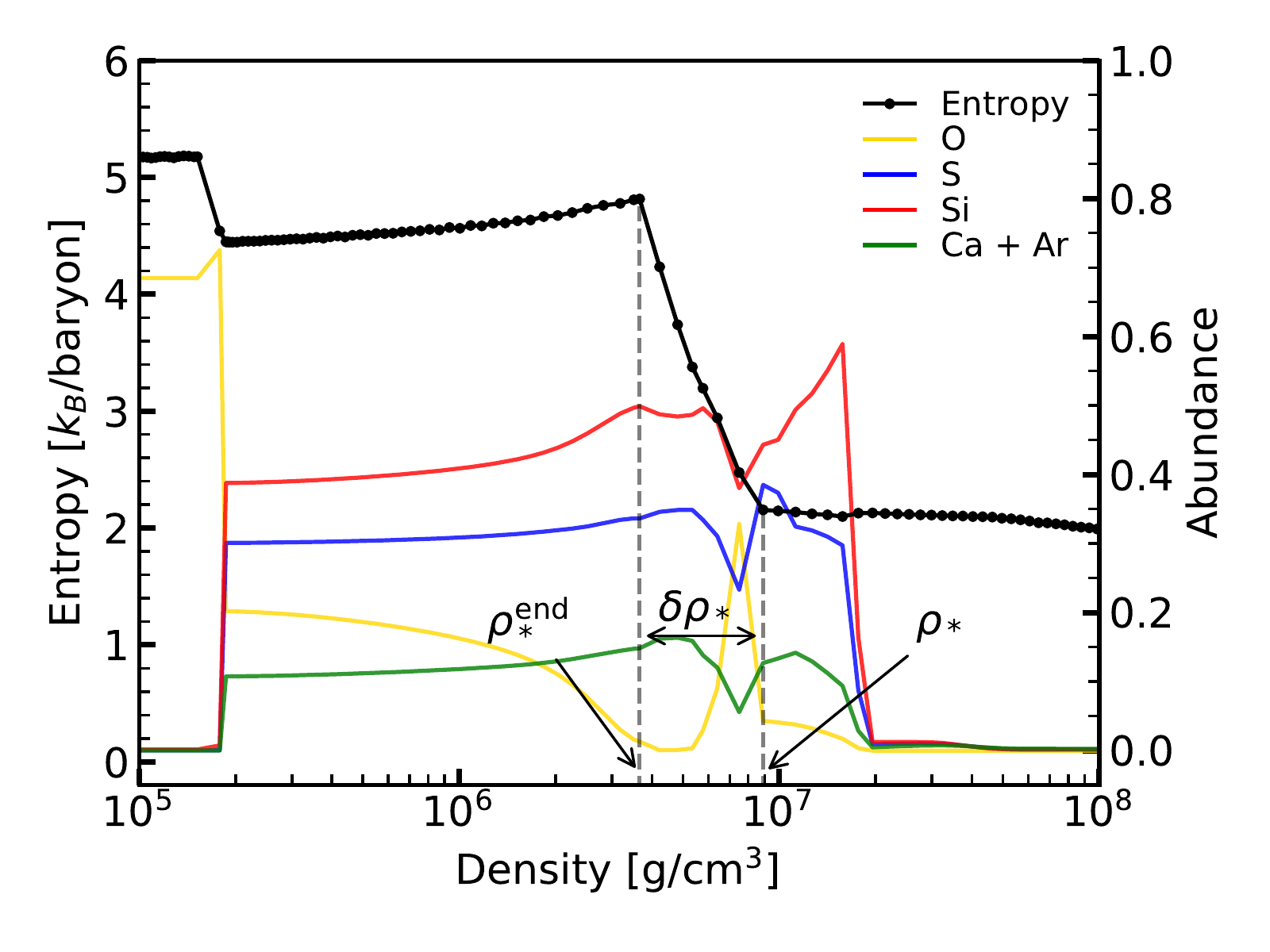}{0.5\textwidth}{(a)}
              \fig{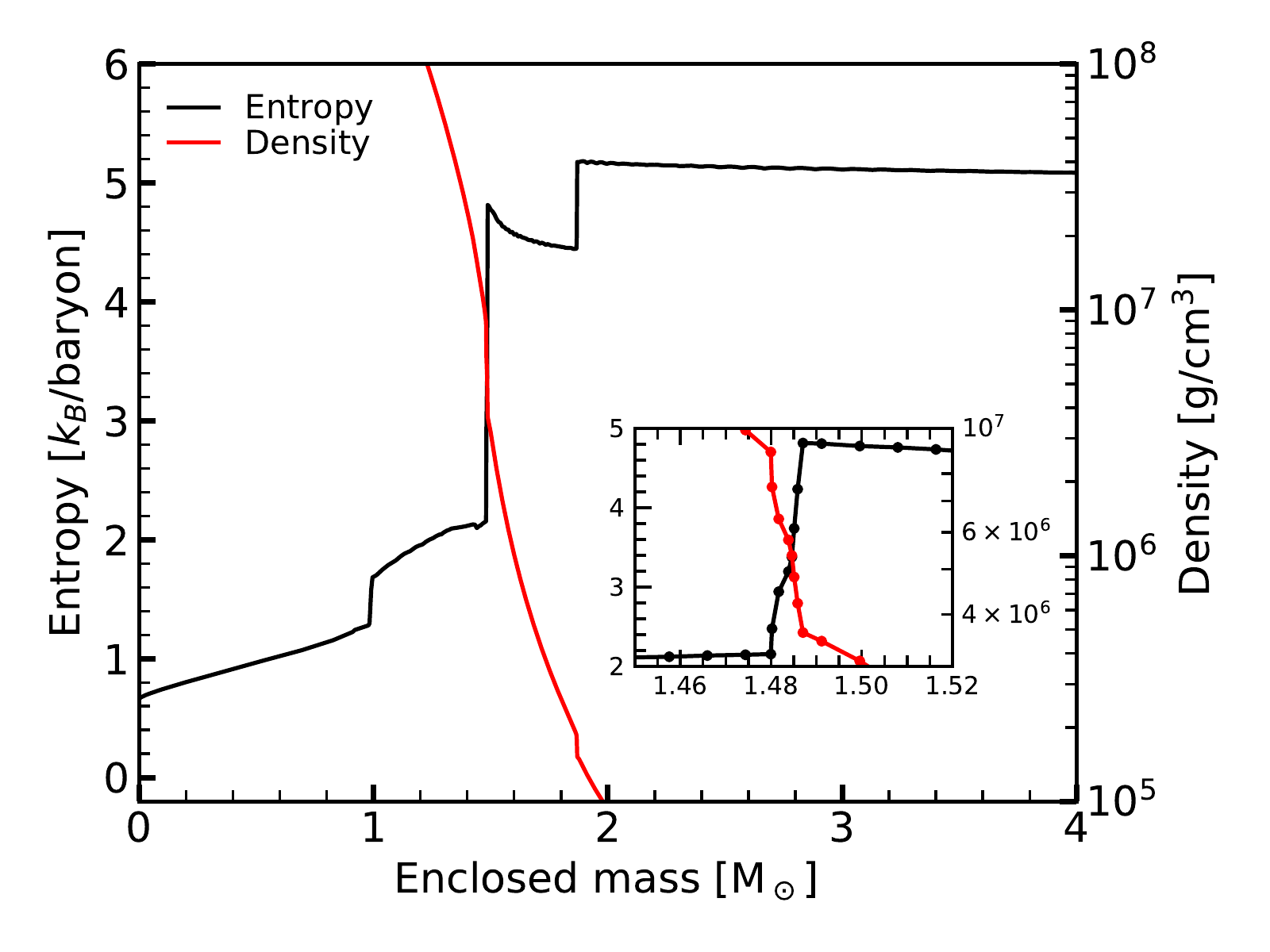}{0.5\textwidth}{(b)}}
    \caption{Example of a typical profile for the 21 M$_\odot$ progenitor from \cite{Sukhbold2016_explodability}. In panel (a) the black solid line shows the entropy as a function of density. The dots correspond to the location of the computational grid. The vertical dashed lines show the zones corresponding to the beginning and the end of the jump, located at densities $\rho_*$ and $\rho_*^{\rm end}$, respectively. There are only a few computational zones along the jump, since both density and entropy vary rapidly with mass and radius, as can be seen in panel (b). We also show some of the most common elements that can be found outside the iron core: oxygen (yellow), sulfur (blue), silicon (red), calcium plus argon (green). In this case, the jump is located inside the silicon shell, and corresponds to the appearance of a pocket of oxygen, whereas the Si-O interface is located at much lower densities, around $\rho = 2 \times 10^5$ g cm$^{-3}$. Panel (b) shows entropy (black) and density (red) as a function of enclosed mass for the same progenitor. The sudden jump corresponding to an increase in entropy and decrease in density happens is located at $\sim$ 1.5 M$_\odot$. A zoom-in of the jump is also shown, where dots correspond to the location of the computational grid. }
    \label{fig:illustration_profile}
\end{figure*}

\begin{figure}
    \centering
    \includegraphics[width=\columnwidth]{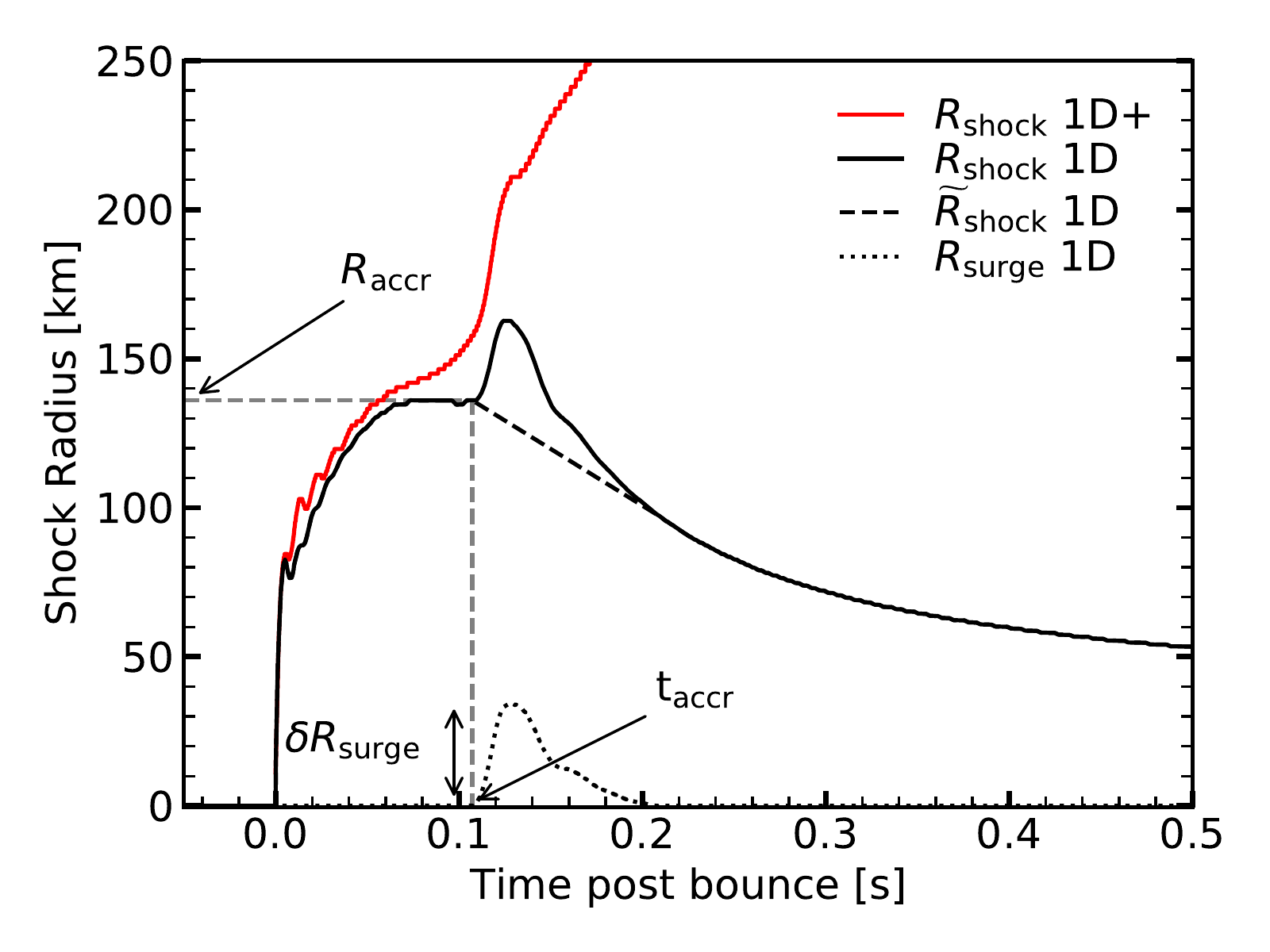}
    \caption{The solid black line shows the shock radius for a standard 1D simulation (i.e. without $\nu$-driven turbulence) of the 21 M$_\odot$ progenitor shown in Figure \ref{fig:illustration_profile}. The solid red line shows the shock radius for a 1D+ simulation (i.e. with $\nu$-driven turbulence) of the same progenitors. The grey dashed lines show the time of accretion $t_{\rm accr}$ and the radius at which the jump is accreted through the shock R$_{\rm accr}$. As one can see, the explosion is triggered right after the accretion of the jump. The black dashed line is a fit that shows the trajectory of the shock if no surge was present. The dotted line is the difference between the solid line and the dashed black lines. The maximum of the dotted line is $R_{\rm surge}$, i.e. how much the shock expands as a consequence of the accretion of the jump. }
    \label{fig:illustration_rsurge}
\end{figure}

\label{sec:density_jump}
Several three-dimensional simulations have shown that progenitors with steep density gradients near the Si/Si-O interface often lead to explosions \citep{Burrows2018_Physical_dependencies_SN,Vartanyan2021_Binary_stars_SiO_interface}, confirming what had been argued in the past. In particular, as explained in Section \ref{sec:previous_studies}, \cite{Ertl2016_explodability} formulated a criterion to predict the outcome of the explosion based on the pre-collapse structure of the progenitor star. The choice of $s=4$ $k_{\rm B}$ baryon$^{-1}$ in their criterion (later applied to a wider set of progenitors by \cite{Sukhbold2016_explodability}) was motivated by the fact that the Si/Si-O interface is generally located at entropies per baryon around that value.

However, \cite{Couch2020_STIR} and \cite{Boccioli2021_STIR_GR} found the explodability as a function of progenitor mass to be very different from the one predicted by \cite{Ertl2016_explodability} and \cite{Sukhbold2016_explodability}. Both of those studies employed \textit{STIR} in 1D simulations with full $\nu$ transport. Hence, it appears that explosions achieved via $\nu$-driven turbulent convection do not obey the explosion criterion proposed by \cite{Ertl2016_explodability}. Since 3D simulations from \cite{Burrows2020_3DFornax} seem to agree with the explodability predicted by \textit{STIR}, we are confident that one can develop a simple explosion criterion based upon the shock dynamics during the accretion of the Si/Si-O interface. 

In spherically symmetric simulations the accretion of the Si/Si-O interface creates a sudden decrease in the ram pressure. This induces a temporary expansion -- or surge -- of the shock. Several semi-analytical models have been used to investigate how fluctuations in pre-shock thermodynamic quantities affect the expansion of the shock \citep{Blondin2003,Nagakura2013_critical_fluct,Nagakura2019_semi_an_prog_asym}. In these studies, it was concluded that these fluctuations can both act as seeds for fluid instabilities that can potentially revive the shock, and also relieve the ram pressure on the shock enough for heating caused by $\nu$-driven turbulence to trigger an explosion. Therefore, in 1D simulations, where $\nu$-driven turbulence is not present, the shock is slowly pushed back down after the initial expansion, continuing its recession towards the PNS. This temporary expansion is almost always caused by the accretion of the Si/Si-O interface, although there are some exceptions.

If the surge is large enough and occurs between $\sim 70$ ms and $\sim 400$ ms after bounce one can expect a 3D simulation to produce an explosion for that progenitor. To quantify the surge we follow the methods outlined in \cite{Schneider2020_EOS_dependence_BH}. Since the surge is connected to a drop in the mass accretion rate, \cite{Schneider2020_EOS_dependence_BH} defined a smooth shock radius $\widetilde{R}_{\rm shock}$ as a fit to the actual shock radius $R_{\rm shock}$ without the inclusion of the region where the accretion rate drops significantly, which is indeed the signature of a surge. 

In our case, we instead define the start of the surge as the time when the density jump (as defined in Section \ref{sec:density_jump}) is accreted through the shock. The two definitions agree very well, which confirms that our definition of the density jump is physically sound since its accretion almost always causes a surge. Moreover, by using our definition we avoid all of the numerical complications derived from having to perform derivatives of the mass accretion rate.\footnote{Sometimes the accretion happens very close to the maximum of the shock radius, where the definition of the surge becomes ambiguous. At early times the mass accretion rate quickly decreases. Therefore, even if it drops as a consequence of the accretion of the jump, the overall trend won't be affected significantly.}

Our fitting procedure is as follows: if the accretion of the density jump happens earlier than 100 ms, we exclude the time window from the accretion until 50 ms after.  We then fit a 2nd-degree polynomial to the shock radius versus time post-bounce. Otherwise, we do a linear fit to the inverse of the shock radius versus time post-bounce, for a time window of 100 ms.\footnote{Early accretions before 100 ms happen around the maximum of the shock radius. Therefore, one expects the shock radius to be roughly quadratically dependent on time. For late-time accretion, one expects $R_{\rm shock} \propto 1/t$ instead.} An example of the fitting procedure, as well as the difference between the simulations with and without \textit{STIR}, is shown in Figure \ref{fig:illustration_rsurge}. The surge is then defined as $R_{\rm surge} = R_{\rm shock} - \widetilde{R}_{\rm shock}$. The maximum of the surge is $\delta R_{\rm surge}$. The larger $\delta R_{\rm surge}$, the more likely is the explosion of that progenitor in realistic 3D simulations. 

The above definition of $R_{\rm surge}$ can only be applied to non-exploding simulations. During an explosion, the shock expands further after the surge, and therefore it is not possible to quantify how much of the expansion is solely due to the accretion of the jump. Hence, in order to quantify $R_{\rm surge}$, we performed a standard 1D simulation for every progenitor, without the inclusion of $\nu$-driven turbulence, in addition to the simulations with \textit{STIR}. To distinguish between these two sets of simulations, we refer to the simple spherically symmetric simulations as ``1D", and to the spherically symmetric simulations with \textit{STIR} as ``1D+" since they incorporate $\nu$-driven convection, a multi-dimensional effect. 

It is worth pointing out that if one uses a loose definition of the density jump (e.g. the location of the $s=4$ $k_{\rm B}$ baryon$^{-1}$ layer), there is a mismatch between when the surge occurs and when that layer is accreted. Therefore, we give a more accurate description of how the jump is defined in the next Section.

\subsection{Definition of the density jump}
As density decreases as a function of enclosed mass, entropy increases. In particular, in layers where the composition changes abruptly, the decrease in density and increase in entropy are very steep, as can be seen in Figure \ref{fig:illustration_profile}. Due to the small range of entropies involved (3-5 $k_{\rm B}$ baryon$^{-1}$) it is more practical to use the entropy to identify the zones that make up the jump. We define the starting zone of a jump as the $i$-th zone such that:

\begin{equation}
    \label{eq:start_jump}
    \frac{s_{i+1} - s_i}{s_i} > 1 \%,
\end{equation}
where s is the entropy per baryon. The end of a jump happens at the $i$-th zone such that:

\begin{equation}
    \label{eq:end_jump}
    \frac{s_{i+1} - s_i}{s_i} < 1 \% \qquad \text{and} \qquad \frac{s_{i+2} - s_{i+1}}{s_{i+1}} < 1 \%.
\end{equation}

The reason for the double condition in Equation \eqref{eq:end_jump} is that sometimes, along the jump, the entropy might not significantly increase in one zone, but keeps increasing after that. Using only the first condition would then underestimate the size of the jump since a fluctuation along the jump would wrongly be identified as the end.

After finding all the jumps in the progenitor, one has to determine which one will cause the surge. If the jump is located at low densities, it will be accreted very late.  Therefore, it will not cause any surge since the shock is already overwhelmed by the ram pressure of the infalling material, and is inevitably receding. On the other hand, if the jump is located at high densities, it will be accreted right after bounce, when the shock is still expanding and neutrino heating is not yet fully developed.  Therefore, it will not cause any surge. We then conservatively chose a range of densities between $9 \times 10^5$ g cm$^{-3}$ and $2 \times 10^7$ g cm$^{-3}$ where the entropy jump can be located. This allows us to include discontinuities that are accreted very early $\sim 60$ ms after bounce, and very late $\sim$ 600 ms after bounce. If no jump is present in this density range, the progenitor will not explode (see Section \ref{sec:criterion} for more details).

If more than one jump is present within this density range (which is often the case), we select the one where the maximum of $\delta \rho^2 / \rho^2$ occurs. Here, $\delta\rho$ is the difference between the density at the start and the end of the jump. The reason for this choice will become clear when we discuss our results in Section \ref{sec:results}. An example of a typical jump in entropy and density is shown in Figure \ref{fig:illustration_profile}. For the remainder of this paper, any quantity ``$q$" calculated at the start (i.e. closer to the center of the star) of the jump will be labeled as $q_*$. If it refers to the end (i.e. closer to the atmosphere of the star) of the jump it will be labeled as $q_*^{\rm end}$. The size of the density jump will be labeled as $\delta\rho_*$.

\section{Results}
\label{sec:results}
To formulate an explosion criterion we selected dynamical properties of the post-bounce phase that correlate with an explosion and then connected them to the properties of the progenitor.
\begin{figure}
    \centering
    \includegraphics[width=\columnwidth]{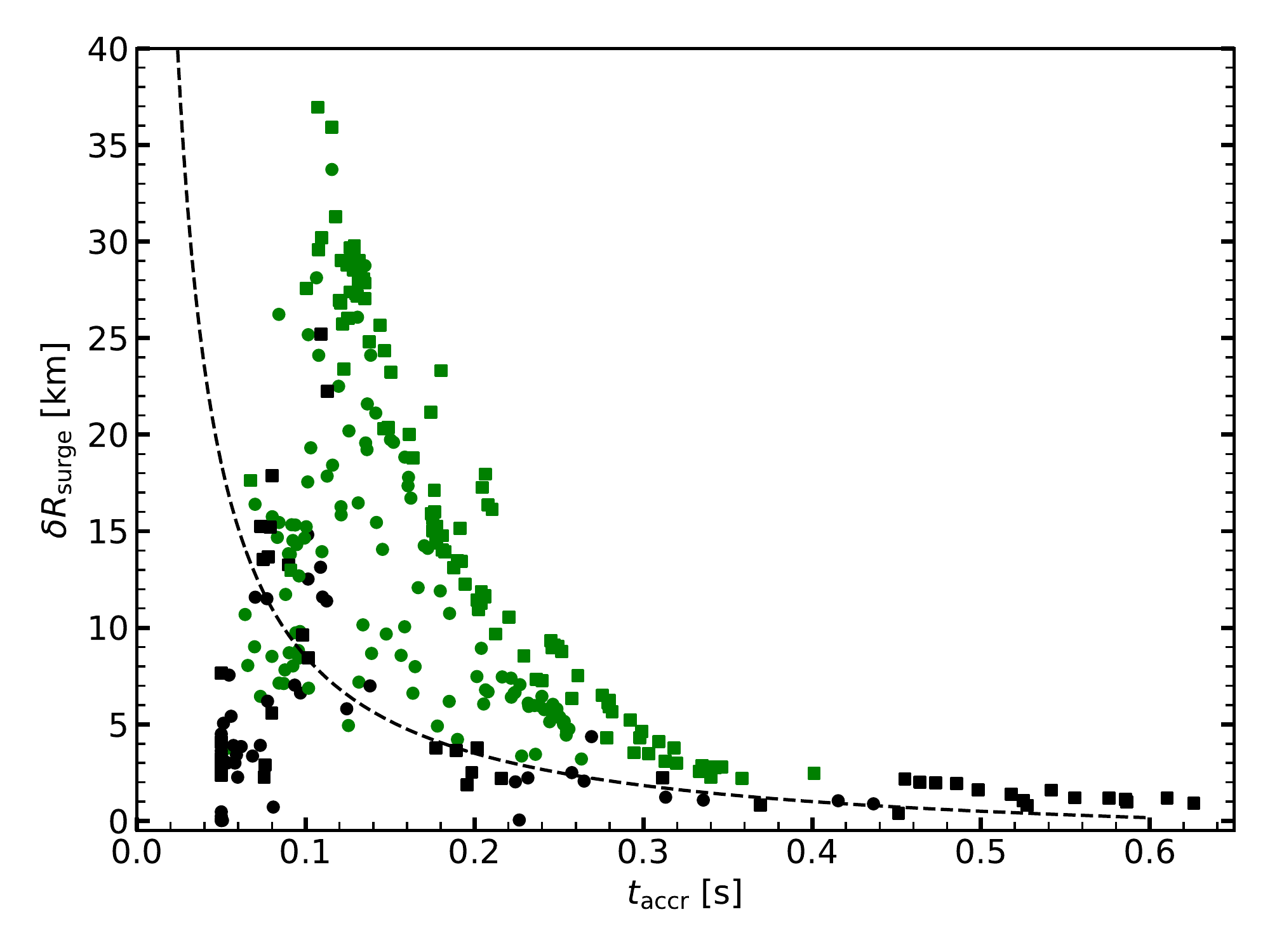}
    \caption{We show how much the shock expands after the accretion of the jump versus the time after bounce when the accretion happens. Each point corresponds to a 1D KEPLER (squares) or FRANEC (circle) simulation. The color of each point indicates that the respective 1D+ simulation with $\alpha_{\rm mLT} = 1.51$ has exploded (green) or resulted in a failed SN (black) The dashed line is $y = -1.5 + 1/x$, and its purpose is simply to guide the eye and show that there is some separation between explosions and failed SN.}
    \label{fig:rsurge_vs_taccr}
\end{figure}

Intuitively, simulations that cause a large $\delta R_{\rm surge}$ should correlate with explosions. At the same time, if the surge happens too early, neutrino-driven convection is not fully developed yet, and therefore there isn't enough heating behind the shock to support further expansion. If the surge happens too late, when the shock has already receded too much, the expansion caused by the accretion of the density jump is not large enough to trigger an explosion. This is summarized in Figure \ref{fig:rsurge_vs_taccr}, where $\delta R_{\rm surge}$ and the time of accretion of the jump $t_{\rm accr}$ are shown for all of our 1D simulations. The dashed line $y = -1.5 + 1/x$, drawn to simply guide the eye, shows that one can separate explosions and failed SN reasonably well. Specifically, surges at early times have to be larger for an explosion to develop, the reason being that neutrino-driven convection is not fully developed yet, and therefore the heating behind the shock is not sufficient to sustain an explosion. The next step is to connect these two quantities to the properties of the progenitor right before collapse. 

A simple argument can be used to quantitatively derive $\delta R_{\rm surge}$. After the initial expansion, the shock stalls at a fixed radius.  This radius then decreases quasi-statically until either an explosion is triggered or a black hole is formed. Therefore, the shock can be approximated as a standing accretion shock with zero net velocity $v_{\rm shk} = 0$. Right before the accretion of the density jump, matter is infalling with a momentum density $\rho_* v_{\rm infall}$. The infall velocity is given by $v_{\rm infall} = \sqrt{2GM / R_{\rm accr}}$ \citep{Janka2001_conditions_shk_revival,Muller2016_prog_connection}, where $R_{\rm accr}$ is the shock radius at the time of accretion and $M$ is the interior mass. After the accretion, the infall velocity has not significantly changed, and therefore the momentum density of the infalling material is $\rho_*^{\rm end} v_{\rm infall}$, with $\rho_* - \rho_*^{\rm end} = \delta \rho_*$. Since momentum has to be conserved, the shock gains a momentum density:

\begin{equation}
    \label{eq:vshk}
    \rho_{\rm shk} v_{\rm shk} = \delta \rho_* v_{\rm infall}.
\end{equation}
Conservation of energy sets the maximum expansion $\delta R_{\rm surge}$ that the shock can experience:

\begin{equation}
    \label{eq:e_cons}
    \frac{1}{2} \rho_{\rm shk} v_{\rm shk}^2 = \rho_{\rm shk} g \delta R_{\rm surge}, 
\end{equation}
where $g = GM/R_{\rm accr}$ is the local gravitational acceleration. Plugging \eqref{eq:vshk} into \eqref{eq:e_cons} and using the expression for $v_{\rm infall}$ given above one finds:
\begin{equation}
    \label{eq:rsurge_vs_rho}
    \frac{\delta R_{\rm surge}}{R_{\rm accr}} = \frac{\delta \rho_*^2}{\rho_{\rm shk}^2} \propto \frac{\delta \rho_*^2}{\rho_*^2}.
\end{equation}

Since $R_{\rm accr} \propto 1/t_{\rm accr}$, one expects the separation found in Fig. \ref{fig:rsurge_vs_taccr} to be characterized by a single number: $\delta R_{\rm surge} / R_{\rm accr}$. We will show later in this Section that this is indeed confirmed by our criterion. However, there is a caveat. Figure \ref{fig:rsurge_vs_taccr} shows that progenitors for which $t_{\rm accr} \gtrsim 0.4$ s result in failed SNe, despite having a non zero $\delta R_{\rm surge}$. This happens because, at such late accretion times, the shock has already receded too much and is inevitably going to fall back onto the PNS. The decrease in ram pressure caused by the accretion of the density jump is therefore not enough to trigger explosions at such late times. 

\begin{figure}
    \centering
    \includegraphics[width=\columnwidth]{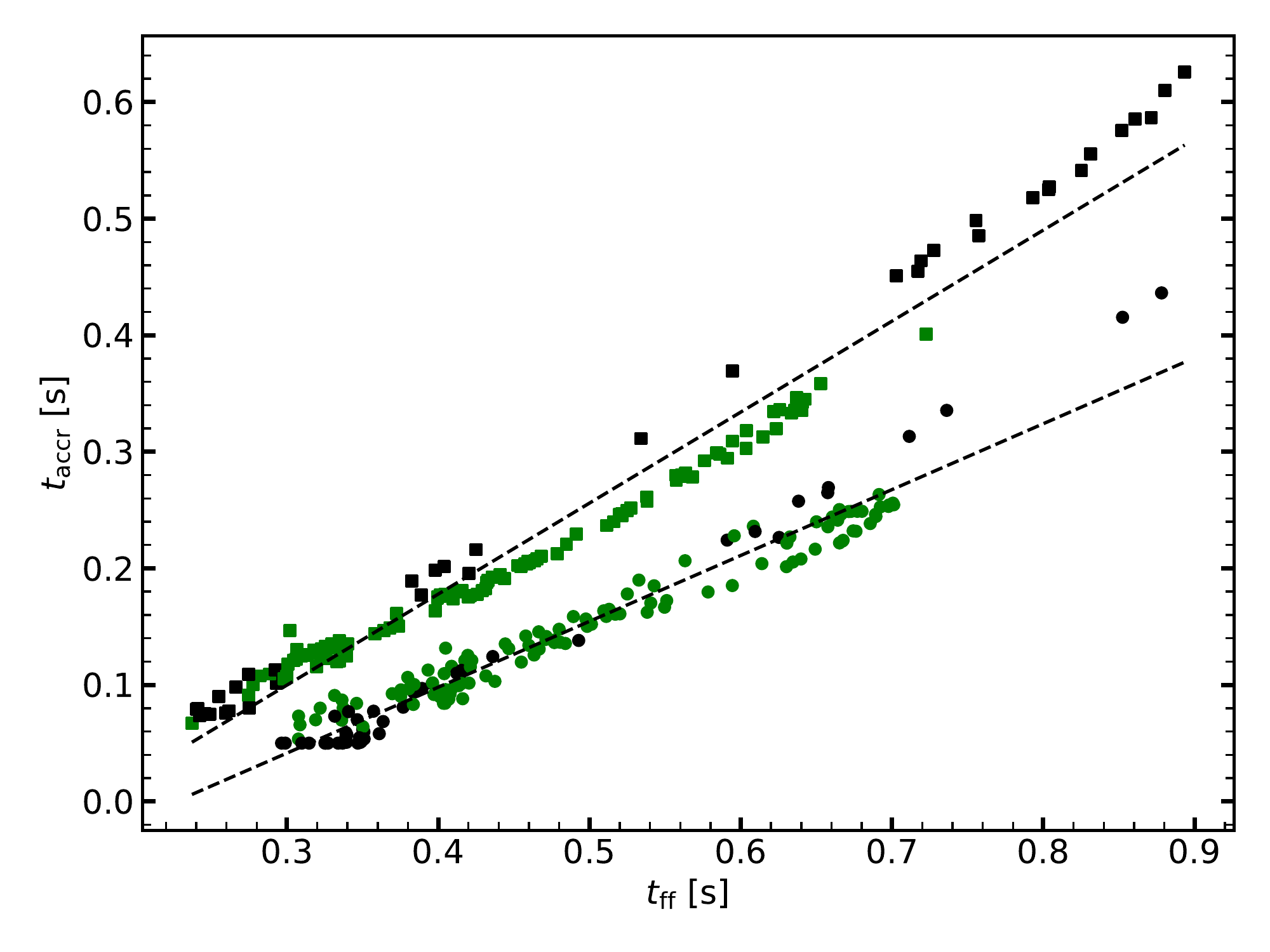}
    \caption{We show the accretion time calculated directly from the simulation versus the accretion time calculated from the pre-collapse progenitor using Equation \eqref{eq:free_fall_time}. Each point represents a different 1D simulation of KEPLER (squares) and FRANEC (circle) progenitors. The color of each point indicates that the respective 1D+ simulation with $\alpha_{\rm mLT} = 1.51$ has exploded (green) or resulted in a failed SN (black). The dashed lines are two separate least-square fits for the KEPLER and FRANEC progenitors, i.e. Equation \eqref{eq:t_accr_profile}. The fit for the KEPLER progenitor yields $t_{\rm accr} = 0.78\times t_{\rm ff} - 0.13$. The fit for the FRANEC progenitor yields $t_{\rm accr} = 0.54\times t_{\rm ff} - 0.12$.}
    \label{fig:taccr_vs_tfrac}
\end{figure}

These progenitors should therefore be treated separately, without the need of performing the simulations to obtain $t_{\rm accr}$. To do that, one has to estimate $t_{\rm accr}$ only from the pre-collapse profile. Therefore, one can calculate the free-fall time of the infalling layer:

\begin{equation}
    \label{eq:free_fall_time}
    t_{\rm ff}= \sqrt{ \pi/4 G \bar{\rho}}, 
\end{equation}
where $\bar{\rho} = M/(4/3\pi r^3)$ is the average density below the infalling layer. The accretion time is a fraction of $t_{\rm ff}$ but is defined with respect to the time of bounce, whereas $t_{\rm ff}$ is defined with respect to the onset of collapse. Taking all of this into account, one can estimate the accretion time using only the pre-collapse density profile, and define $\widetilde{t}_{\rm accr}$ as:

\begin{equation}
    \label{eq:t_accr_profile}
    \widetilde{t}_{\rm accr} = C t_{\rm ff} - t_0 = C \sqrt{ \frac{\pi}{4 G \bar{\rho}}} - t_0~~,
\end{equation}
where $C$ is a constant smaller than 1, and t$_0$ is the time of bounce. It's worth noting that the time of bounce will be different for each progenitor, but the spread is relatively small, as can be inferred from Figure \ref{fig:taccr_vs_tfrac}. Therefore, we consider $t_0$ as a single representative value for all the times of bounce. 

Depending on when the simulation starts (i.e. how close to bounce) these constants will be different. Specifically, in this paper, we use progenitor stars from both KEPLER (a hydrodynamic code) and FRANEC (a hydrostatic code). Therefore, the onset of collapse will be at different pre-bounce times, which will change $C$ and $t_0$ in the above expression. Consequently, we fit KEPLER and FRANEC progenitors separately. 

Using $t_{\rm accr}$ from our 1D simulations, we fit Equation \eqref{eq:t_accr_profile} and find $(C,t_0)\rvert_{\rm KEPLER} = (0.78,0.13$ s) and $(C,t_0)\rvert_{\rm FRANEC} = (0.54,0.11$ s), as shown in Figure \ref{fig:taccr_vs_tfrac}. $C$ is a dimensionless quantity, whereas $t_0$ is in units of seconds. 

\subsection{Explosion criterion}
\label{sec:criterion}
To determine whether a progenitor explodes or not, we use the 1D+ simulations with $\alpha_{\rm MLT} = 1.51$, as explained in Section \ref{sec:calibration}. We can now formulate two explosion criteria: criterion (a) based upon dynamical properties ($\delta R_{\rm surge}$/$R_{\rm accr}$ and $t_{\rm accr}$); and criterion (b) based upon pre-collapse properties ($\delta \rho_*^2/\rho_*^2$ and $\widetilde{t}_{\rm accr}$).

The first part of our criteria reads:
\begin{enumerate}[label=\alph*)]
    \item if $t_{\rm accr} > 0.4$ s the star will not explode. Otherwise, the star \emph{can} explode, based on the discussion that follows.
    \item if $\widetilde{t}_{\rm accr} > 0.4$ s the star will not explode. Otherwise, the star \emph{can} explode, based on the discussion that follows.
\end{enumerate}

This leads to the prediction that that 17 progenitors do not explode. Of these, only the 22.8 M$_\odot$ KEPLER progenitor explodes despite having $\widetilde{t}_{\rm accr} = 0.430$ s and $t_{\rm accr} = 0.401$ s, and is therefore misclassified by both criteria. After determining that progenitors with $t_{\rm accr} > 0.4$ s ($\widetilde{t}_{\rm accr} > 0.4$ s) do not explode, we exclude them from the subsequent discussion. Moreover, we also exclude the progenitors that do not have any density jump satisfying the definition given in section \ref{sec:density_jump}. None of these progenitors explode, as expected. 

\begin{figure}
    \centering
    \includegraphics[width=\linewidth]{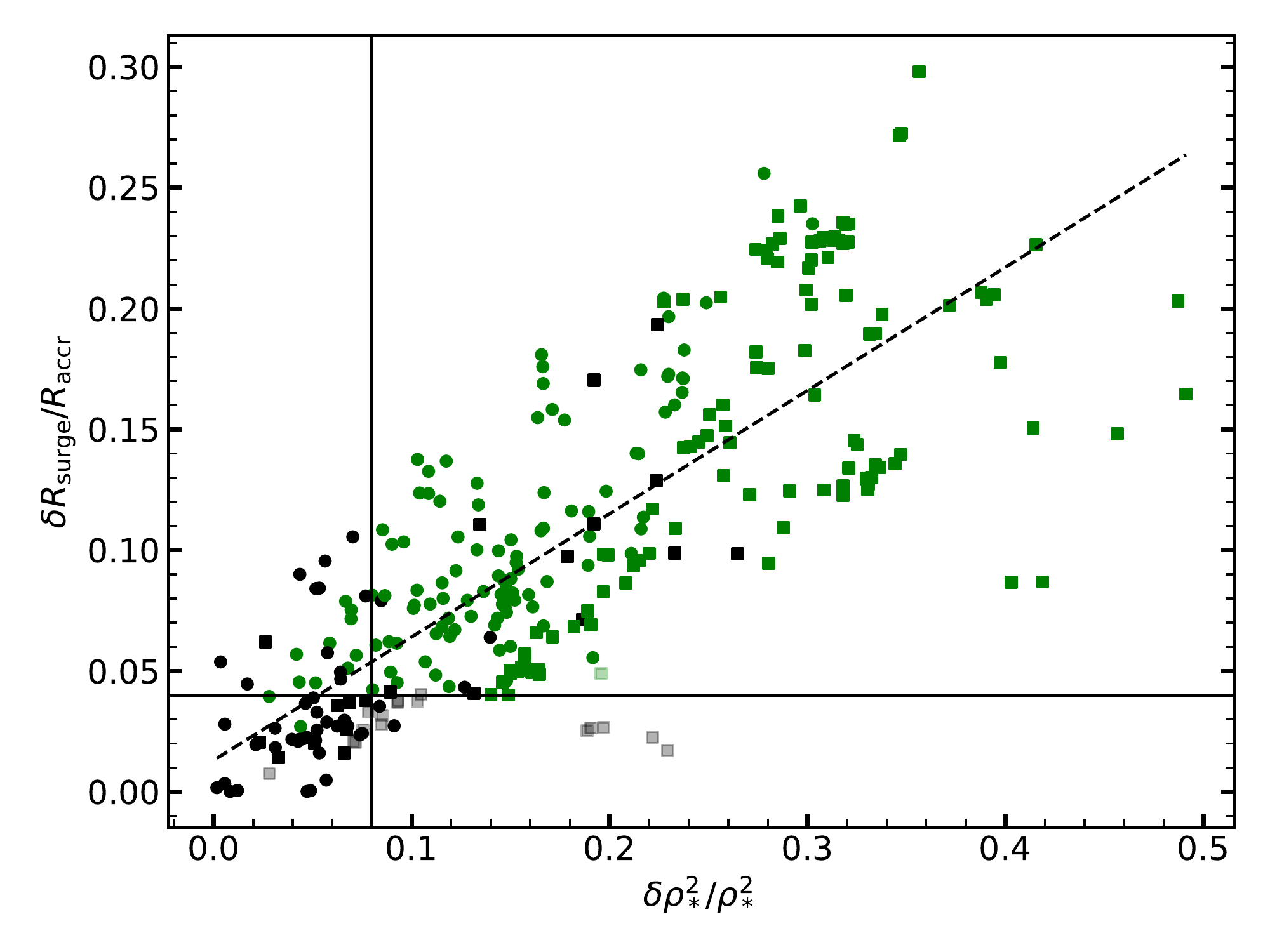}
    \caption{Each point corresponds to either a KEPLER (squares) or FRANEC (circle) progenitor. The quantity on the y-axis has been calculated using 1D simulations, whereas the quantity on the x-axis has been calculated from the pre-collapse density profile with the procedure outlined in Section \ref{sec:density_jump}. The color of each point indicates that the respective 1D+ simulation with $\alpha_{\rm mLT} = 1.51$ has exploded (green) or resulted in a failed SN (black). The dashed line shows the least-square fit $y = 0.51x + 0.013$. The horizontal line is located at $\delta R_{\rm surge}$/$R_{\rm accr} = 0.04$ and divides explosions (above) from failed SN (below) according to criterion (a). The vertical line is located at $\delta \rho_*^2/\rho_*^2 = 0.08$ and divides explosions (right) from failed SN (left) according to criterion (b).}
    \label{fig:rsurge_vs_rho}
\end{figure}

Based on the discussion above, the remaining progenitors should follow Equation \ref{eq:rsurge_vs_rho}, and as can be seen from Figure \ref{fig:rsurge_vs_rho} that is indeed the case. For completeness, progenitors with $\widetilde{t}_{\rm accr} > 0.4$ s are shown as shaded symbols, but they are not included in the analysis. It is interesting to notice that some of them have very large $\delta \rho_*^2/\rho_*^2$ but small $\delta R_{\rm surge}$/$R_{\rm accr}$. However, for accretions at very late times, it is very hard to estimate the surge correctly, and therefore these variations are likely to be a consequence of numerical noise. 

Interestingly, on the right side of Figure \ref{fig:rsurge_vs_rho}, there are some progenitors with $\delta R_{\rm surge}$/$R_{\rm accr} > 0.25$ and others with $\delta \rho_*^2/\rho_*^2 > 0.4$ that seem to significantly deviate from the best-fit line, shown as a dashed black line. We verified that in the case of the progenitors with $\delta R_{\rm surge}$/$R_{\rm accr} > 0.25$, $\delta R_{\rm surge}$ has been overestimated by our fitting algorithm. This also partially explains the vertical spread around the best-fit line of all progenitors, since as discussed in previous sections the estimation of $\delta R_{\rm surge}$ suffers from numerical noise. Analogously, for the progenitors with $\delta \rho_*^2/\rho_*^2 > 0.4$, $\delta R_{\rm surge}$ has been underestimated.

We find that our fitting procedure is relatively simple, physically justified, and has an overall good performance. Therefore, we believe that fine-tuning it to properly account for these outliers defeats the purpose of finding a simple criterion that can connect dynamical and pre-collapse properties. Moreover, none of these outliers are misclassified, and therefore changing the fitting procedure would not change the results. The only change would be a better correlation between $\delta R_{\rm surge}$/$R_{\rm accr}$ and $\delta \rho_*^2/\rho_*^2$, which we only use as a consistency check for Equation \ref{eq:rsurge_vs_rho}, but it does not enter in our criterion.

We can now formulate the second part of our explosion criteria:
\begin{enumerate}[label=\alph*)]
    \item If $\delta R_{\rm surge}$/$R_{\rm accr} > 0.04$ the progenitor explodes. Otherwise, it results in a failed SN.
    \item If $\delta \rho_*^2/\rho_*^2 > 0.08$ the progenitor explodes. Otherwise, it results in a failed SN.
\end{enumerate}

Criterion (a), formulated using dynamical properties, produced $\sim 9 \%$  false positives and $\sim 1 \%$  false negatives. Some of these are located near the horizontal line in Figure \ref{fig:rsurge_vs_rho}, which represents $\delta R_{\rm surge}$/$R_{\rm accr} = 0.04$, and can be considered statistical fluctuations. The imbalance between false positives and negatives is partially due to the definition of $\delta R_{\rm surge}$, which can be affected by numerical noise and lead to an overestimation of $\delta R_{\rm surge}$, especially when the accretion happens close to the maximum of the shock radius. This generates a significant number of false positives on the top left quadrant of Figure \ref{fig:rsurge_vs_rho}, that are however correctly classified by criterion (b). A few more false positives can be found on the top right quadrant of Figure \ref{fig:rsurge_vs_rho}, and are quite far from the horizontal line. All of them are KEPLER progenitors and, as explained later in the section, some of them explode at slightly larger values of $\alpha_{\rm MLT}$. They are also misclassified by criterion (b).

Criterion (b), formulated using pre-collapse properties, produced $\sim 5 \%$  false positives and $\sim 4 \%$  false negatives. Some of these are located near the vertical line in Figure \ref{fig:rsurge_vs_rho}, which represents $\delta \rho_*^2/\rho_*^2 = 0.08$, and can be considered statistical fluctuations. However, some misclassifications are quite far from the dividing line and deserve an explanation. Like in criterion (a), some misclassifications happen on the top-left quadrant, but in this case, these are false negatives, i.e. progenitors with $\delta \rho_*^2/\rho_*^2 < 0.08$ that nevertheless explode. These are all FRANEC progenitors, and the ones with the smallest values of $\delta \rho_*^2/\rho_*^2$ have all very low compactnesses $\xi_{\rm 2.5} < 0.05$.

Because of their low compactness, these progenitors are characterized by very steep density profiles and low mass accretion rates. Therefore, the ram pressure exerted on the shock is small.  This means that even density discontinuities accreted very late can be sufficient to trigger an explosion. Moreover, they generally tend to be easier to explode, and some of them can indeed explode very early without the need for the accretion of a density jump. A more nuanced criterion is likely needed to describe these progenitors, but this goes beyond the purposes of this paper and will be addressed in future work. Finally, several misclassifications of KEPLER progenitors happen in the top right quadrant, as already described for criterion (a). A more detailed analysis of why such misclassifications arise is given in section \ref{sec:criterion_vs_alpha}.

The analysis presented so far was done on the combined set of progenitors. However, it is interesting to apply the criteria to the KEPLER and FRANEC sets separately. The results are reported in Table \ref{tab:criterion_table}. Criterion (a) performs equally well on both sets, with no remarkable differences. This is a consequence of the overestimation of $\delta R_{\rm surge}$ that occurs in some cases due to uncertainties in the fitting algorithm.

Criterion (b), instead, has significantly different performances on the two sets. The false positives are much higher for the KEPLER progenitors. This can be attributed to the outliers on the top-right quadrant in Figure \ref{fig:rsurge_vs_rho} which, as discussed in Section \ref{sec:criterion_vs_alpha}, can be accounted for by tweaking $\alpha_{\rm MLT}$. By doing that, the false positives drop from 7.1 \% to 2.6 \%, which is much closer to the value of 3.1 \% found for FRANEC progenitors.

The false negatives are instead much higher for FRANEC progenitors. This can be attributed to the finer resolution of FRANEC progenitors with masses $M < 13$ M$_\odot$, compared to KEPLER. These progenitors have very low compactness and, as described above, are likely to explode even in absence of the accretion of a strong density jump. If one excludes from the analysis progenitors with $\xi_{2.5} < 0.05$, the false negatives for FRANEC progenitors drop from 6.9 \% to 3.3 \%. This indicates that low compactness progenitors can explode even without the presence of a strong density jump. 

Interestingly, KEPLER progenitors with $\xi_{2.5} < 0.05$ obey the criterion since the 12.5 M$_\odot$ and 12.75 M$_\odot$ do not explode, contrary to the expectation that low compactness progenitors should be more susceptible to explosions. Therefore, one can only surmise that a more nuanced criterion is needed to account for the behavior of low compactness progenitors, but this goes beyond the scope of this paper.

\begin{table}
\centering 
    \begin{tabular}{l|c|c|c}
        \toprule
        \toprule
            Criterion (a) & Combined & FRANEC  & KEPLER  \\
        \midrule
        \midrule
            False positives &  8.5 \% & 8.8 \% & 8.4 \% \\
            False negatives & 1.0 \% & 1.2 \% & 0.6 \% \\
        \midrule
            Total  &  9.5 \% & 10.0 \% & 9.0 \% \\
        \midrule 
        \midrule 
            \multicolumn{4}{c}{} \\
        \midrule 
        \midrule 
            Criterion (b) & Combined & FRANEC  & KEPLER  \\
        \midrule  
        \midrule  
            False positives &  5.0 \% & 3.1 \% & 7.1 \% \\
            False negatives &  3.8 \% & 6.9 \% & 0.7 \% \\
        \midrule
            Total  &  8.8 \% & 10.0 \% & 7.8 \% \\
        \bottomrule
        \bottomrule
    \end{tabular}
\caption{Performance of criterion (a) and (b) for both FRANEC and KEPLER progenitor sets combined, as well as separately. The total misclassification percentage is reported, broken down into false positives and negatives.}
\label{tab:criterion_table}
\end{table}

\subsection{Comparison with Wang (2022)}
A similar study was very recently carried out by \cite{Wang2022_prog_study_ram_pressure}, who used 100 2D simulations to calibrate their criterion instead of our 1D+ approach. Moreover, they only applied this criterion to KEPLER progenitors, whereas we also included FRANEC models. 

The first difference is in the fact that \cite{Wang2022_prog_study_ram_pressure} use the ram pressure $P_{\rm ram}$ instead of density as the main quantity in their criterion. However, the two are interchangeable, since $P_{\rm ram} = \rho v_{\rm infall}^2$, and since the infall velocity $v_{\rm infall}$ does not change when the jump is accreted one finds that $\delta P_{\rm ram} / P_{\rm ram} \approx \delta \rho / \rho$. 

The selection of the density jump of \cite{Wang2022_prog_study_ram_pressure} is very similar to the one we adopted, described in section \ref{sec:density_jump}. In our case, we look for the maximum of $\delta \rho^2 / \rho^2$ in the density range between $9 \times 10^5$ g cm$^{-3}$ and $2 \times 10^7$ g cm$^{-3}$, whereas \cite{Wang2022_prog_study_ram_pressure} look for the maximum of $\delta P_{\rm ram} / P_{\rm ram}$ in the region between the outer envelops and the Si-O interface. 

\begin{figure*}
    \centering
    \gridline{\fig{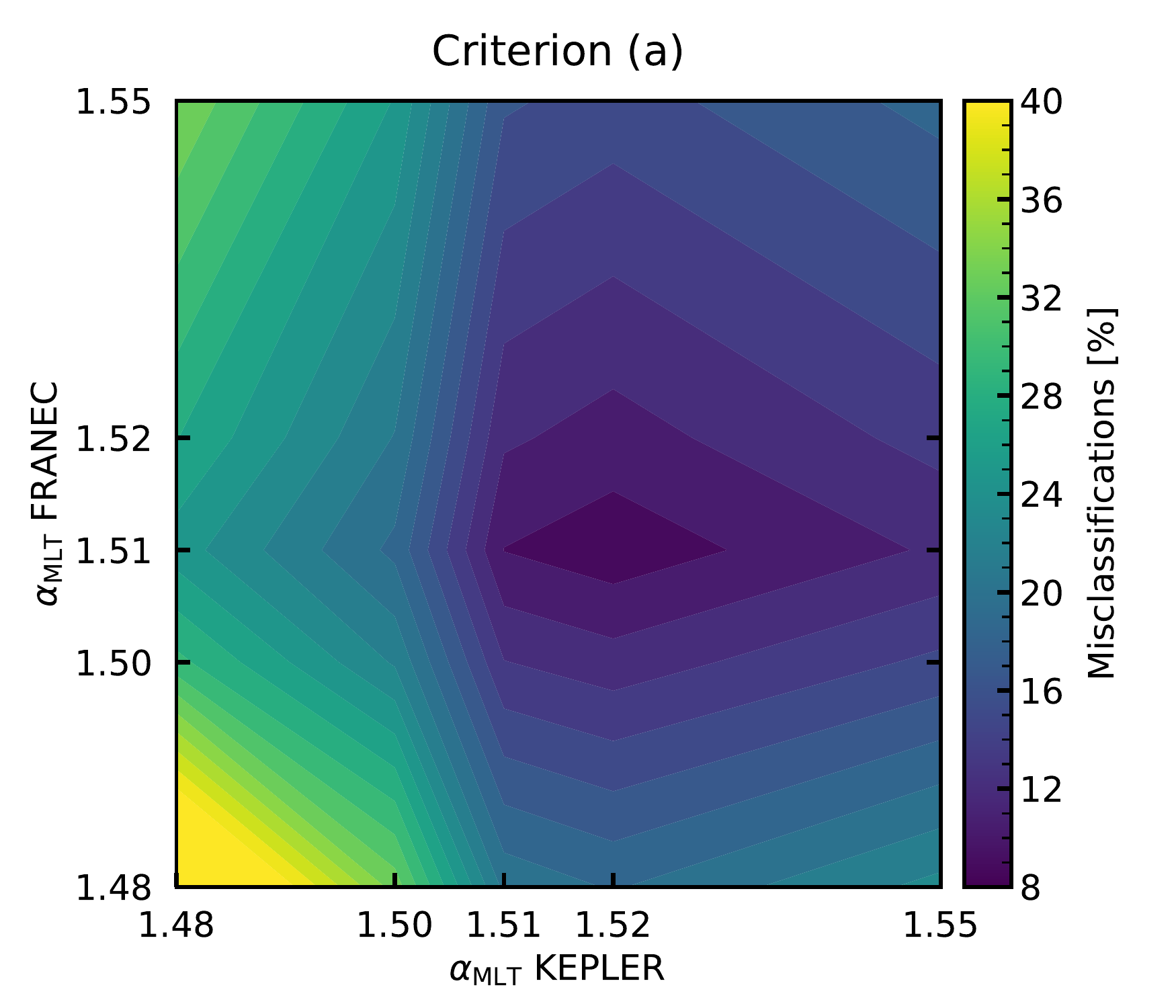}{0.5\textwidth}{(a)}
              \fig{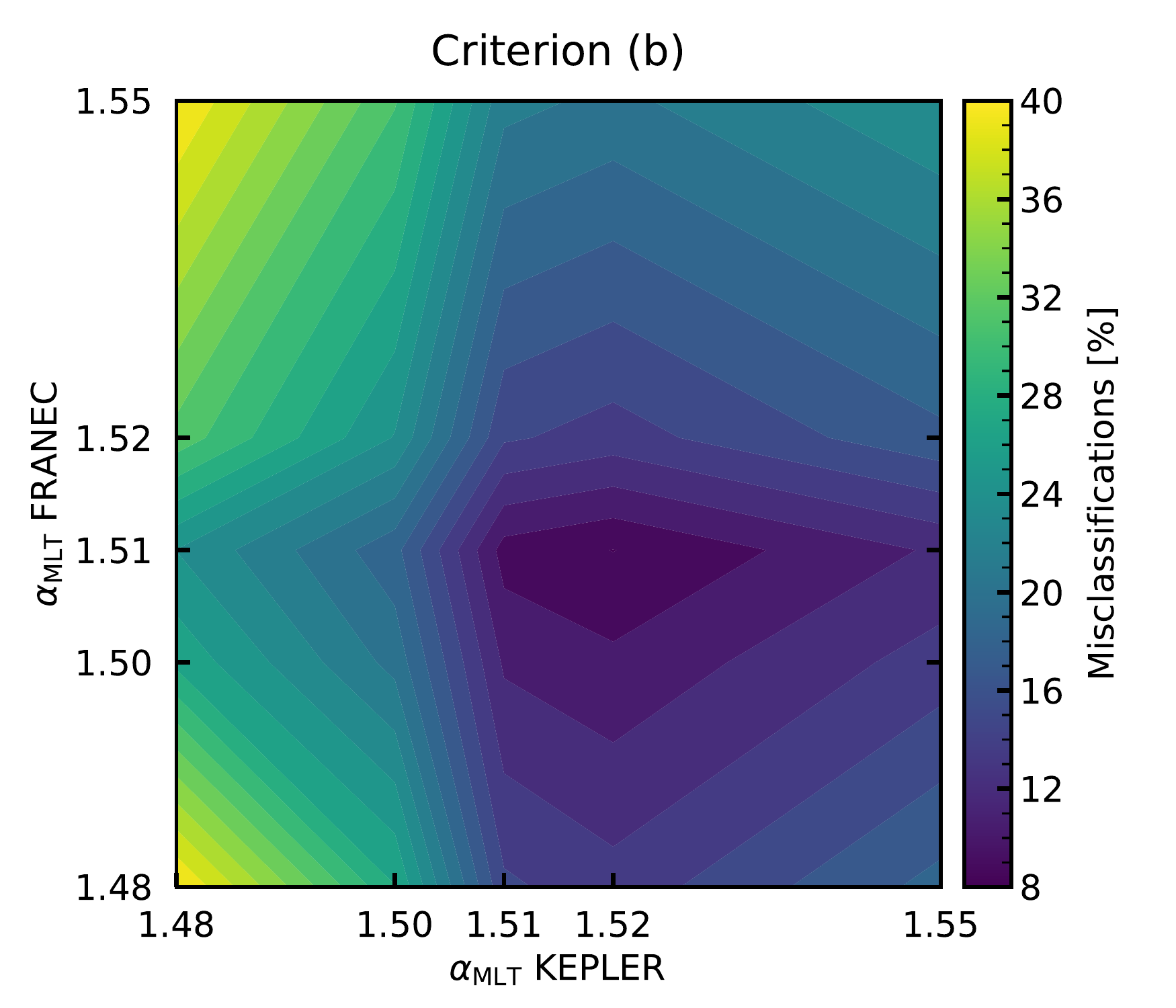}{0.5\textwidth}{(b)}}
    \caption{Explodability criteria as a function of $\alpha_{\rm MLT}$. Panel (a) shows criterion (a), i.e. the one formulated with $\delta R_{\rm surge}/R_{\rm accr}$ and $t_{\rm accr}$. Panel (b) shows criterion (b), i.e. the one formulated with $\delta\rho_*/\rho_*$ and $\widetilde{t}_{\rm accr}$. The y-axis shows $\alpha_{\rm MLT}$ used for the FRANEC models, while the x-axis shows $\alpha_{\rm MLT}$ used for the KEPLER models. The color bar shows how many misclassifications the criterion yields. Notice that KEPLER and FRANEC progenitors don't necessarily have to be run with the same value of $\alpha_{\rm MLT}$.  However, the best performance for both criteria is around $\alpha_{\rm MLT}$ = 1.51 for both sets of progenitors }
    \label{fig:criterion_vs_alpha}
\end{figure*}

In their paper, \cite{Wang2022_prog_study_ram_pressure} define the Si-O interface as ``\textit{the density discontinuity closest to the inner boundary of the oxygen shell in which the oxygen abundance is above 15 percent}". Therefore, with this definition, the Si-O interface is sometimes well inside the silicon shell. For example, for the profile shown in Figure \ref{fig:illustration_profile}, the definition of the jump by \cite{Wang2022_prog_study_ram_pressure} agrees with ours, and the density jump is entirely inside the silicon shell. For that reason, in this paper, we have used the expression ``Si/Si-O interface" rather than ``Si-O interface" since the drop in density is not necessarily located at the bottom of the oxygen layer.

Quantitatively, our criterion agrees extremely well with the one by \cite{Wang2022_prog_study_ram_pressure}. In their case, they predict explosions for progenitors where ${\rm max}(\delta P_{\rm ram} / P_{\rm ram}) > 0.28$. Since $\delta P_{\rm ram} / P_{\rm ram} \approx \delta \rho / \rho$, this condition becomes ${\rm max}\left(\delta\rho^2/\rho^2\right) > 0.078$, which is almost exactly the same as our $\delta\rho_*^2/\rho_*^2 > 0.08$. The small discrepancy is due to slight differences in the definition of the density jump. \cite{Wang2022_prog_study_ram_pressure} define $\delta P_{\rm ram} = P_{\rm ram}(t + \delta t) - P_{\rm ram}(t)$, where $\delta t = 10$ ms and $t$ is calculated using Equation \eqref{eq:free_fall_time}. Basically, instead of calculating the density jump directly from the pre-collapse progenitor, as we do, they use the density profile to calculate the accretion time and look at density variations in 10 ms windows. Therefore, the actual values they obtain might be slightly different from ours, but the agreement is still excellent.

We can therefore conclude that this is a robust criterion to predict the explodability of massive stars. In their criterion, \cite{Wang2022_prog_study_ram_pressure} use 2D and 3D simulations computed with \texttt{Fornax} to determine the explodability of stars. In our criterion, we use 1D+ simulations computed with \texttt{GR1D}, a completely different code where $\nu$-driven turbulence is added through a time-dependent MLT model. This serves as further confirmation that the main ingredient missing from 1D simulations is $\nu$-driven turbulence, and that STIR does a good job simulating it in spherical symmetry.

\subsection{Dependency of the explosion criteria on $\alpha_{\rm MLT}$}
\label{sec:criterion_vs_alpha}
To understand the origin of the top-right quadrant misclassification, i.e. progenitors resulting in failed SNe despite being expected to explode according to both criterion (a) and (b), it's useful to analyze how the criteria change as a function of $\alpha_{\rm MLT}$. In this analysis, we do not require that FRANEC and KEPLER progenitors are simulated using the same value of $\alpha_{\rm MLT}$. The percentage of misclassifications is shown in Figure \ref{fig:criterion_vs_alpha}. Even without forcing the two progenitor sets to have the same $\alpha_{\rm MLT}$, both criteria give the best predictions in the range $1.5 \leq \alpha_{\rm MLT} \leq 1.52$ for both KEPLER and FRANEC models. 

The same best-fit range was derived in section \ref{sec:calibration} based on the calibrations of \cite{Boccioli2022_EOS_effect}, as well as on the total explosion fraction shown in Figure \ref{fig:explosion_fraction}. This serves as further confirmation of the robustness and consistency of STIR. A more detailed analysis of Figure \ref{fig:criterion_vs_alpha} shows that the best performance of both criteria is obtained with $\alpha_{\rm MLT} = 1.51$ for FRANEC progenitors and $\alpha_{\rm MLT} = 1.52$  for KEPLER progenitors. This partially explains the misclassifications on the top right quadrant of Figure \ref{fig:rsurge_vs_rho}. Two of those KEPLER progenitors explode at $\alpha_{\rm MLT} = 1.52$, and three more have $\tau_{\rm adv} / \tau_{\rm heat} > 1.1$, which means that they are very close to an explosion. Only two progenitors are then left unaccounted for in the top-right quadrant.

By fine-tuning $\alpha_{\rm MLT}$ and allowing $\alpha_{\rm MLT} = 1.51$ for FRANEC progenitors and $\alpha_{\rm MLT} = 1.52$ for KEPLER progenitors, one can improve both criteria. Additionally, one can go a step further and consider successful explosions even the 1D+ KEPLER simulations where the shock is not revived, but where $\tau_{\rm adv} / \tau_{\rm heat} > 1.1$. The latter condition is based on the concept that even if the 1D+ simulation does not explode, a slightly more efficient $\nu$-driven convection would yield a successful explosion, because of the large $\tau_{\rm adv} / \tau_{\rm heat}$. That would improve both criteria, yielding a 7.5 \% misclassification rate for criterion (a) and a 7 \% misclassification rate for criterion (b).

This criterion was however developed based on very general principles. Therefore, too much fine-tuning does not add anything to the actual physical interpretation of the criteria, even though it explains the origin of a few KEPLER outliers. The more significant result is that both criteria yield very low misclassification rates even when using the same value of $\alpha_{\rm MLT} = 1.51$. Therefore, it can be inferred that STIR does not depend on the pre-collapse history of the progenitor, but simply on the post-bounce thermodynamic conditions in the gain region, as expected.

\subsection{Comparison with Ertl (2016)}
\label{sec:ertl_comparison}
The present study was partially motivated by a mismatch between the explodability found by \cite{Ertl2016_explodability} and the one found by \cite{Couch2020_STIR} and \cite{Boccioli2021_STIR_GR}. It is therefore useful to understand where the discrepancy comes from. In this section, we address this difference by focusing only on the KEPLER progenitors from \cite{Sukhbold2016_explodability} since they were used in \cite{Ertl2016_explodability}. 

In the simulations of \cite{Ertl2016_explodability} the explosion was triggered using the engine from \cite{Ugliano2012}, which was calibrated from observations (i.e. explosion energy and ejected nickel mass) of core-collapse supernovae. In their model, they replace the PNS with an inner boundary from which neutrinos are emitted. The luminosity of the emitted neutrinos is calculated based on the mass accretion rate as well as on the thermodynamic properties of the infalling material. The inner boundary's contraction follows an analytical prescription fitted to reproduce the explosion energy and ejected nickel mass of SN 1987a \citep{Sonneborn1987a}, assuming a 20 M$_\odot$ progenitor. The luminosities in these models tend to be overestimated by a factor of $\sim 2-3$ compared to realistic 3D simulations.

\begin{figure}
    \centering
    \includegraphics[width=\columnwidth]{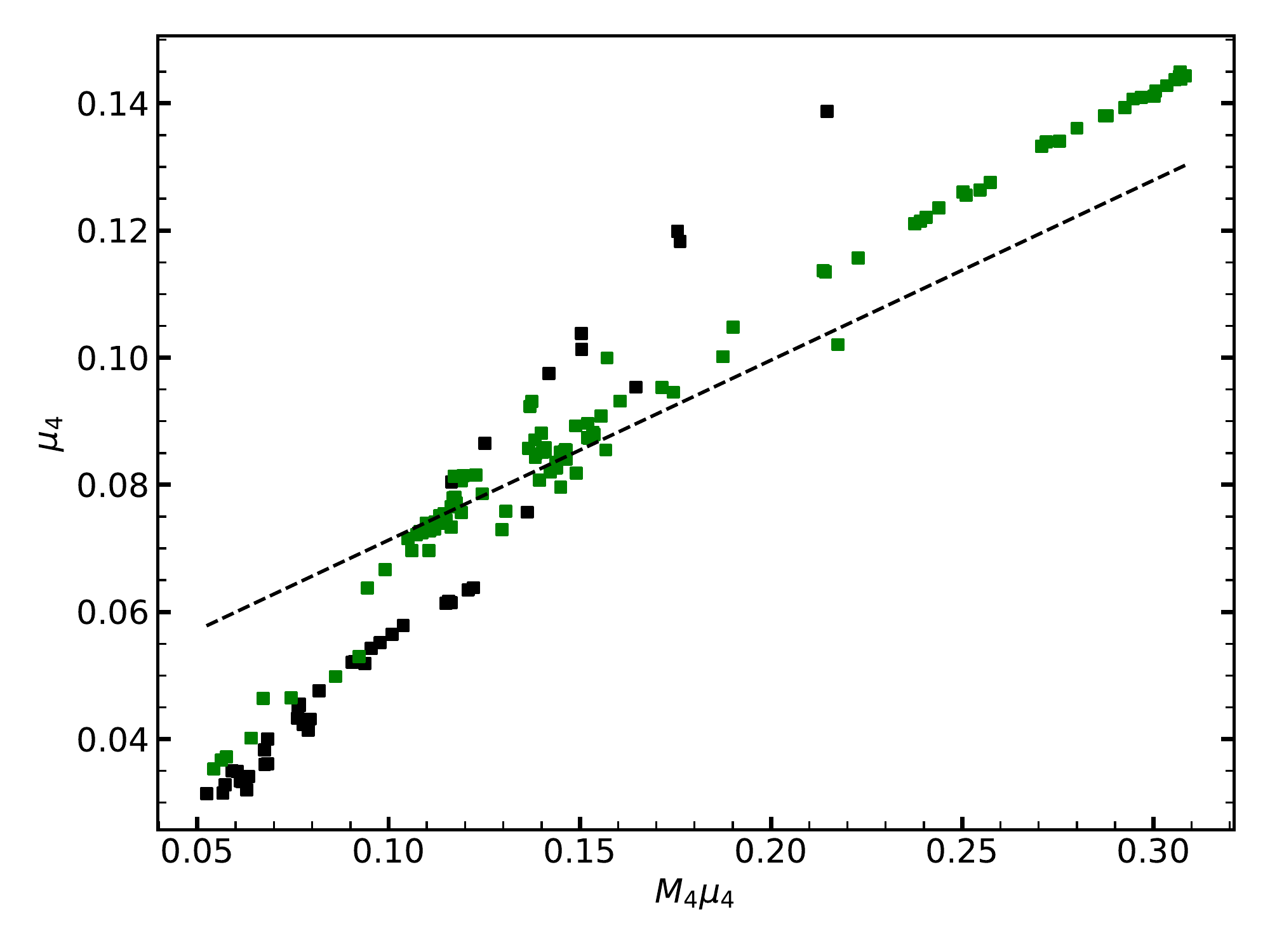}
    \caption{This is the plane defined by \cite{Ertl2016_explodability}, where $M_4$ is the mass location of the $s=4$ k$_{\rm B}$ baryon$^{-1}$ layer, and $\mu_4$ is a measure of the mass gradient and is defined as $\frac{0.3 {\rm M}_\odot \times 1000 {\rm km}}{r(M_4 + 0.3 {\rm M}_\odot) - r(M_4)}$. The dashed line represents the condition relative to the N20 engine of \cite{Ertl2016_explodability}. The squares are the KEPLER progenitors from \cite{Sukhbold2016_explodability}, colored in green if they explode, according to our 1D+ simulation, and black in they don't. It is expected that progenitors resulting in explosions (failed SN) lie below (above) the line. This is not the case for our simulations}
    \label{fig:Ertl_criterion}
\end{figure}

Firstly, by looking at Figure \ref{fig:Ertl_criterion}, it's very clear that our predicted explosions do not follow the criterion from \cite{Ertl2016_explodability}. They find that there is a line in the $\mu_4-M_4\mu_4$ plane that separates explosions (below the line) from failed SN (above the line). We find almost the exact opposite, although no line can separate our predicted explosions from failed SN. The reason behind the mismatch is that large $\mu_4$ typically occur for progenitors with a large $\delta\rho_* / \rho_*$, which explode according to our criterion. A similar mismatch was found in \cite{Couch2020_STIR}, whose Figure 13 is very similar to our Figure \ref{fig:Ertl_criterion}. Since they also use STIR to trigger explosions in 1D, this does not come as a surprise. Instead, it serves as further evidence that what is causing this mismatch is the inclusion of $\nu$-driven turbulence via STIR.

\begin{figure*}
    \centering
    \gridline{\fig{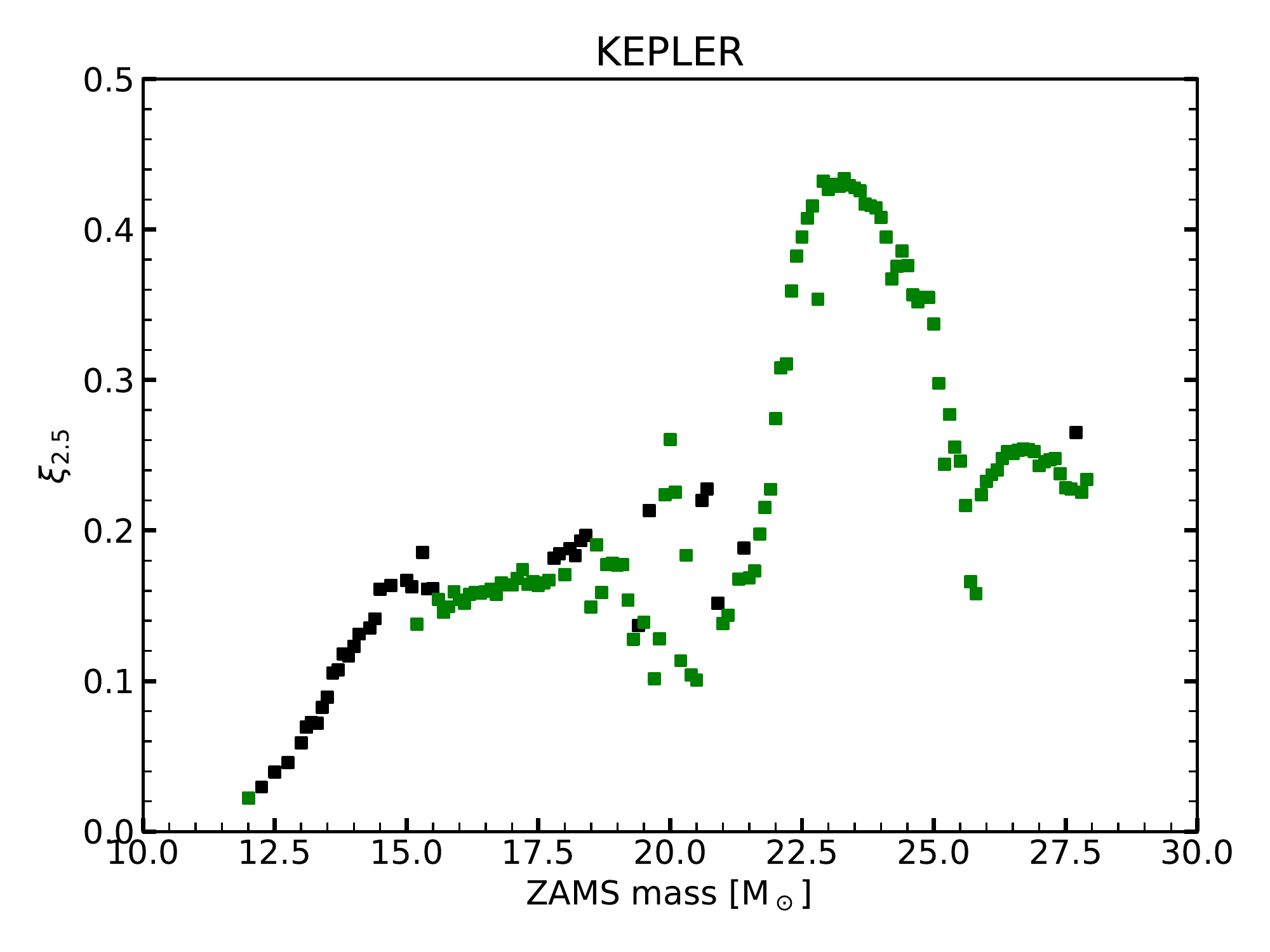}{0.5\textwidth}{(a)}
              \fig{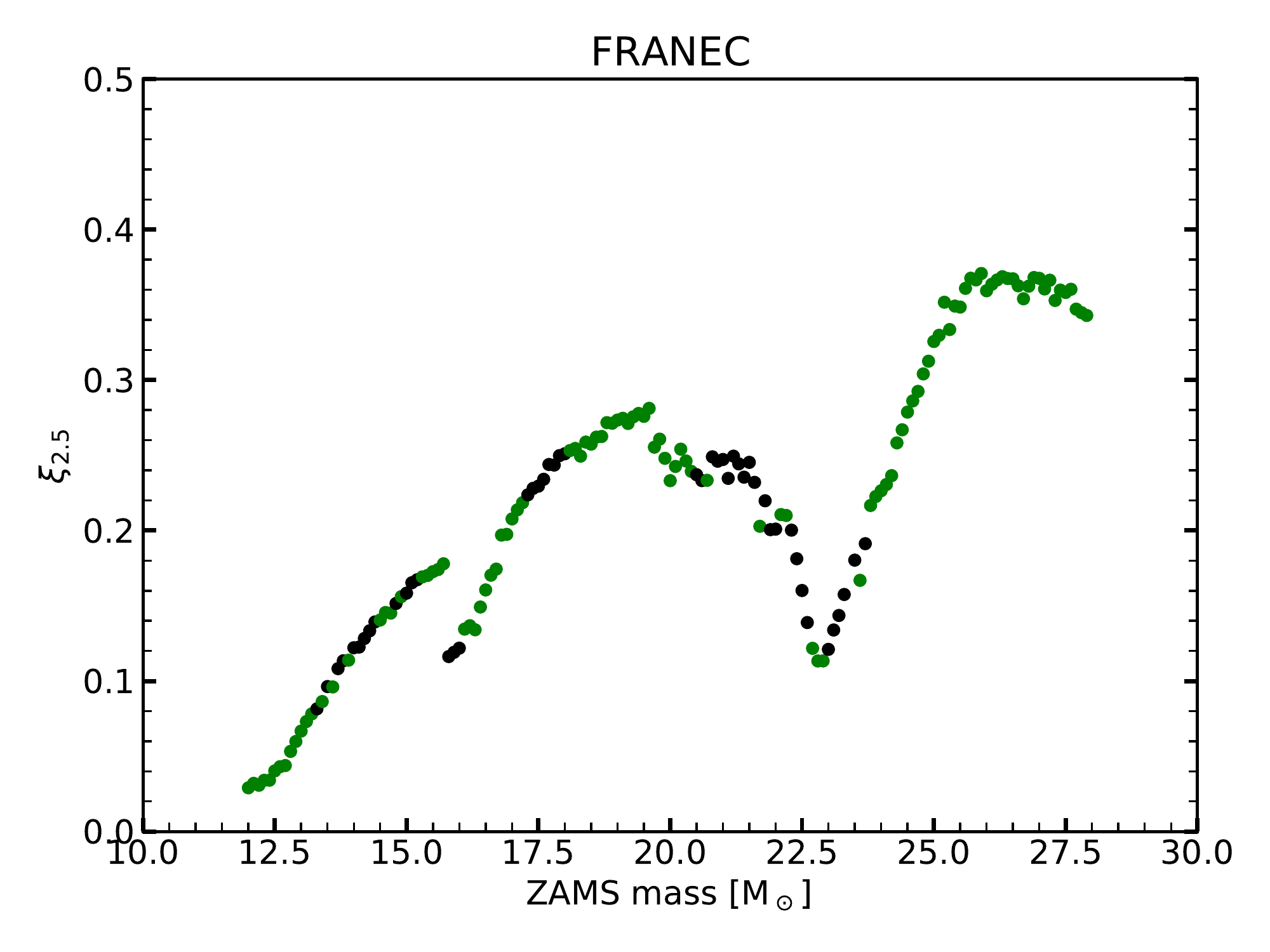}{0.5\textwidth}{(b)}}
    \caption{Compactness $\xi_{2.5}$ for the KEPLER progenitors from \cite{Sukhbold2016_explodability} on the left panel and FRANEC progenitors from \cite{Chieffi2020_presupernova_models} on the right panel, as a function of ZAMS mass. The color of each point indicates that the respective 1D+ simulation with $\alpha_{\rm mLT} = 1.51$ has exploded (green) or resulted in a failed SN (black). There is no correlation between explodability and $\xi_{2.5}$ or ZAMS mass. Moreover, it should be noted that for $\xi_{2.5} > 0.3$ both panels predict explosions, whereas \cite{Ertl2016_explodability} and \cite{Muller2016_prog_connection} predict failed SNe. The reason is that $\nu$-driven turbulence in our simulations is particularly efficient. }
    \label{fig:comp_vs_ZAMS_mass}
\end{figure*}

The bottom two panels of Figure 7 from \cite{Ertl2016_explodability} show that progenitors that have large mass accretion rates right after the infall of the layer with $s=4$ result in failed supernovae.  This suggests that in their explosion models this is the most important feature. Indeed, to trigger the explosion one needs a small ram pressure ahead of the shock, a condition that is satisfied by small mass accretion rates. The results that they find are in line with the critical luminosity condition criterion. However, their luminosities are much larger than what is seen in 3D simulations, and most importantly their results suggest that all progenitors with mass accretion rates above a certain value lead to failed supernovae, regardless of the luminosity. 

In our supernova simulations, it is $\nu$-driven turbulence that provides extra pressure behind the shock, and therefore even large ram pressures can be overcome if the neutrino heating (aided by turbulence) is large enough. An example of this can be seen in the progenitors with masses $22 < M < 25$ M$_\odot$.  These have the largest mass accretion rates amongst the progenitors simulated, as a consequence of their large compactness $\xi_{2.5} > 0.3$. According to the criterion by \cite{Ertl2016_explodability} these progenitors should not explode, whereas we find the opposite. For these progenitors, the mass accretion rates are counterbalanced by large neutrino heating.  This is only possible with the inclusion of $\nu$-driven convection.

This is also evident in Figure \ref{fig:comp_vs_ZAMS_mass}, to be compared with Figure 6 from \cite{Muller2016_prog_connection}, who find qualitatively similar results to \cite{Ertl2016_explodability}. For completeness, we show both the KEPLER progenitors on the left panel and the FRANEC progenitors on the right panel. It is clear that even progenitors with large $\xi_{2.5}$, predicted to fail by a simple criterion based on compactness, can successfully explode. Furthermore, no clear explodability trend can be seen in either the ZAMS mass or the compactness. 

\begin{figure*}
    \centering
    \gridline{\fig{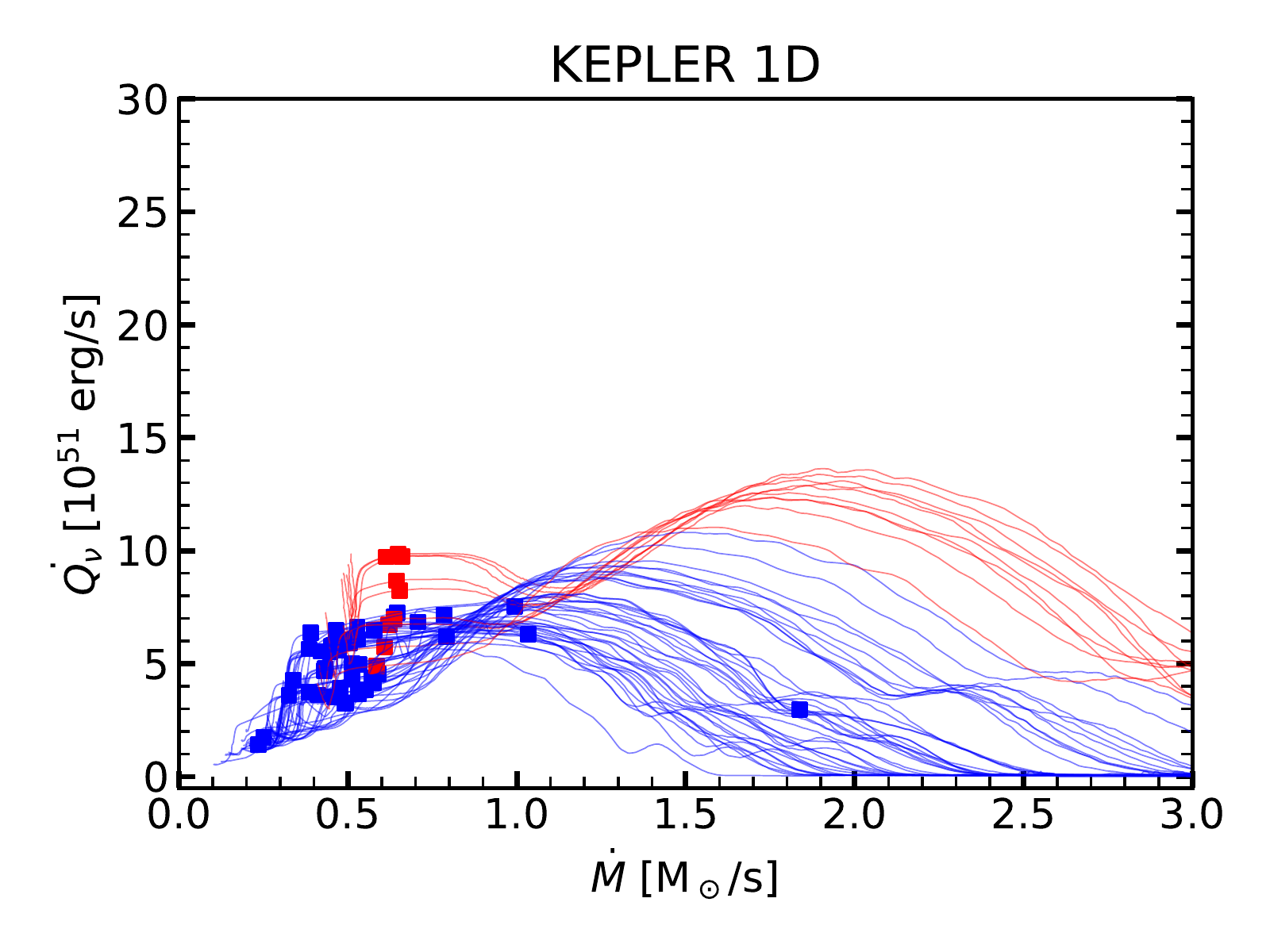}{0.5\textwidth}{(c)}
              \fig{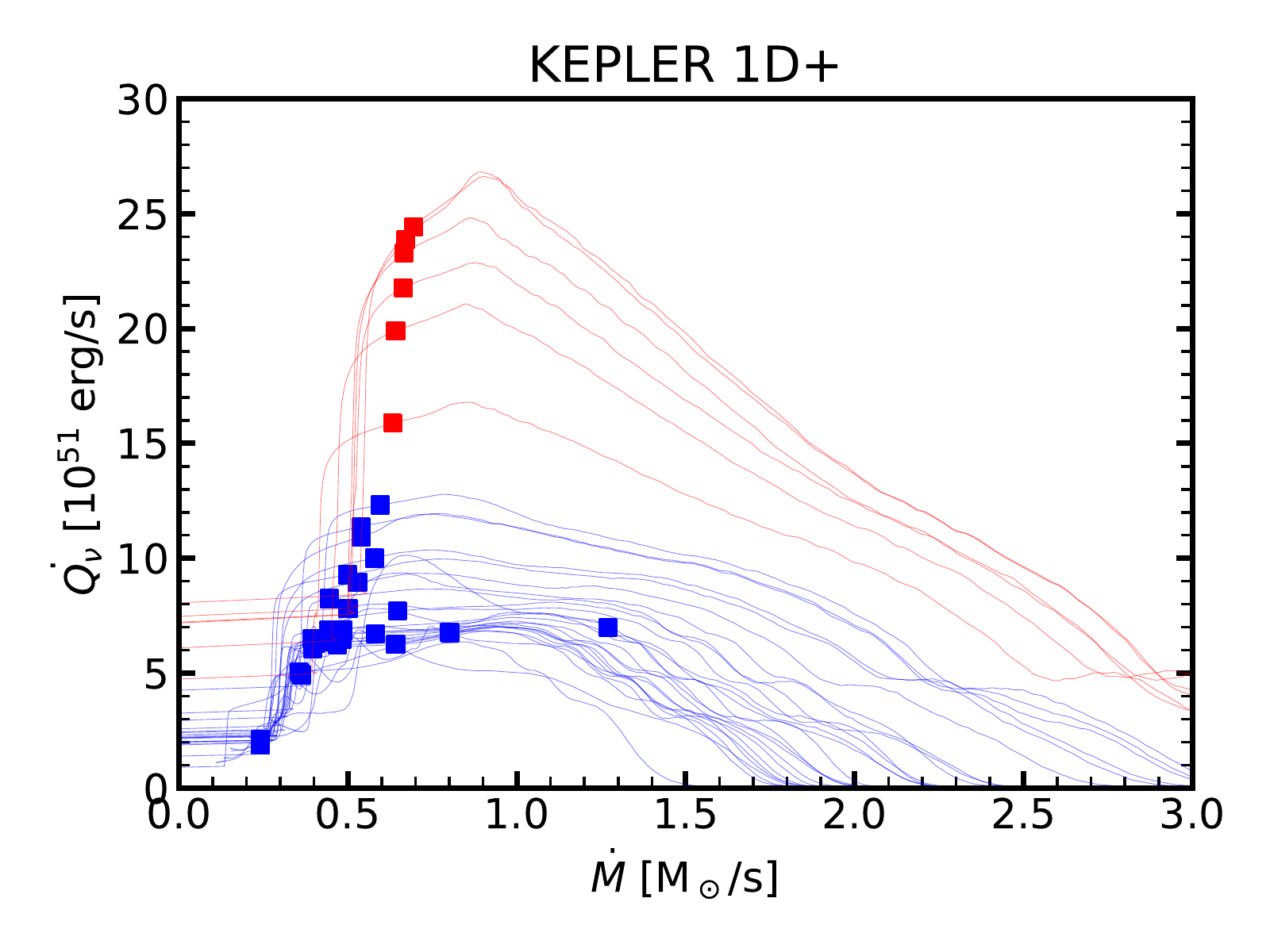}{0.5\textwidth}{(d)}}
    \caption{In all plots only KEPLER progenitors are shown. The same plots for FRANEC progenitors are shown in Figure \ref{fig:maccr_heat_KEPLER}. Red lines correspond to progenitors with $\xi_{2.5} > 0.3$, whereas blue lines correspond to all the others. Panel (a) shows the mass accretion rate $\dot{M}$ calculated at 500 km as a function of time after bounce for 1D+ simulations. Panel (b) shows the net neutrino heating $\dot{Q}_\nu$ in the gain region as a function of time after bounce for the same 1D+ simulations. The bottom panels only show a subset of the simulated progenitors, to avoid cluttering. Panel (a) shows the evolution track in the $\dot{M}-\dot{Q}$ plane for the 1D simulations for selected progenitors. The square points mark the time of accretion of the density jump $t_{\rm accr}$. Panel (b) is the same as panel (a) but for 1D+ simulations. Time goes from right to left.  }
    \label{fig:maccr_heat_KEPLER}
\end{figure*}

The mass range $22 < M < 25$ M$_\odot$ coincides with the largest compactness progenitors in the KEPLER set. Again, despite being predicted to fail by the criterion of \cite{Ertl2016_explodability}, these models explode. We emphasize that it is $\nu$-driven convection that generates the heating necessary for these progenitors to sustain shock expansion after the accretion of the density jump. This can be inferred by the $\dot{M}-\dot{Q}_\nu$ plane in Figure \ref{fig:maccr_heat_KEPLER}. Panel (a) shows the tracks of a few 1D models (i.e. without \textit{STIR}), whereas panel (b) shows the tracks for the same 1D+ models.

The reason we chose the $\dot{M}-\dot{Q}_\nu$ plane rather than $\dot{M}-L_\nu$ is that the latter does not show the influence of $\nu$-driven convection. The luminosity is determined by the thermodynamic properties of the PNS, whereas the contribution from $\nu$-driven convection is only present in the gain region, and therefore cannot be captured by the $\dot{M}-L_\nu$ plane. 

\begin{figure*}
    \centering
    \gridline{\fig{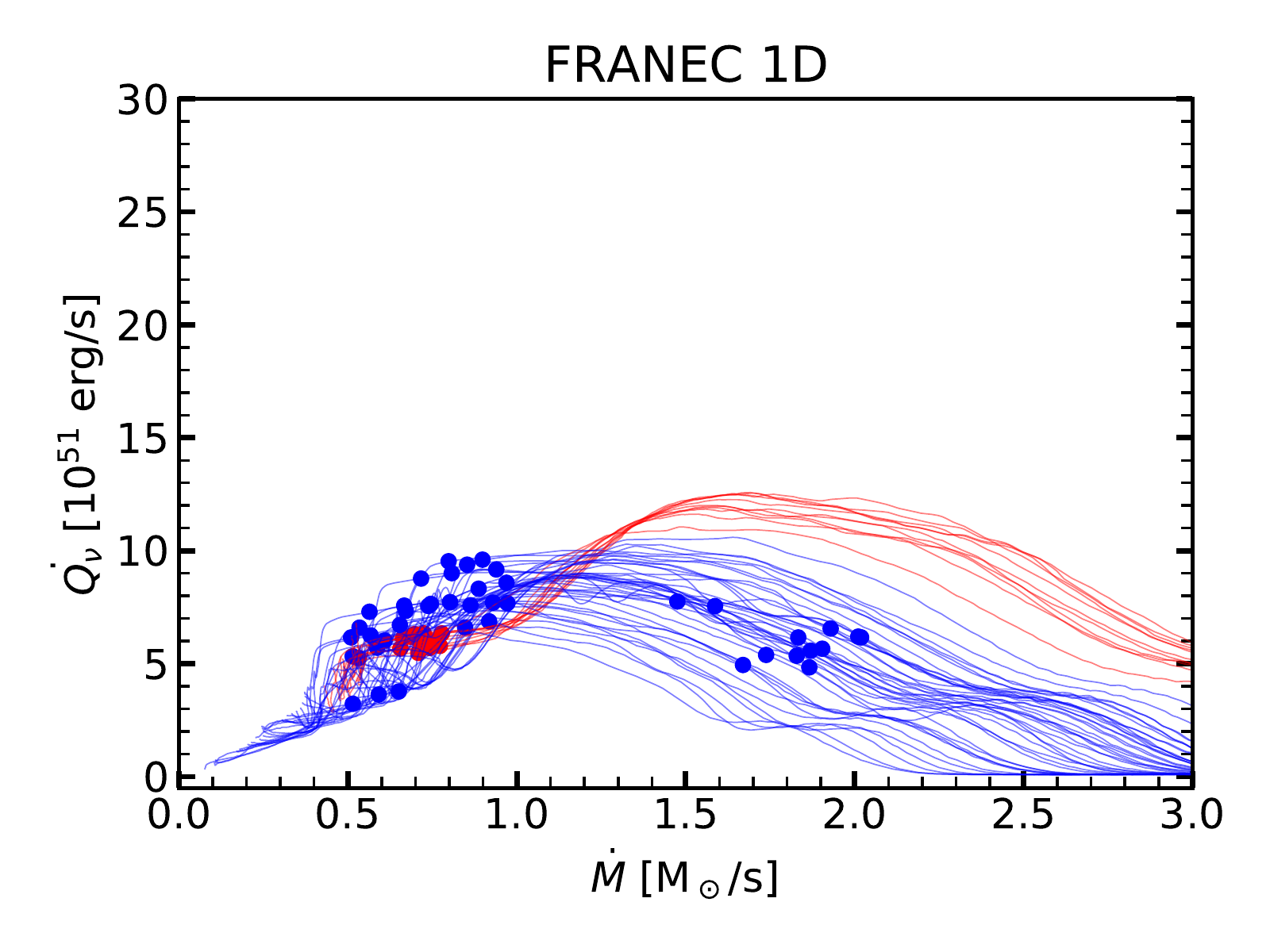}{0.5\textwidth}{(c)}
              \fig{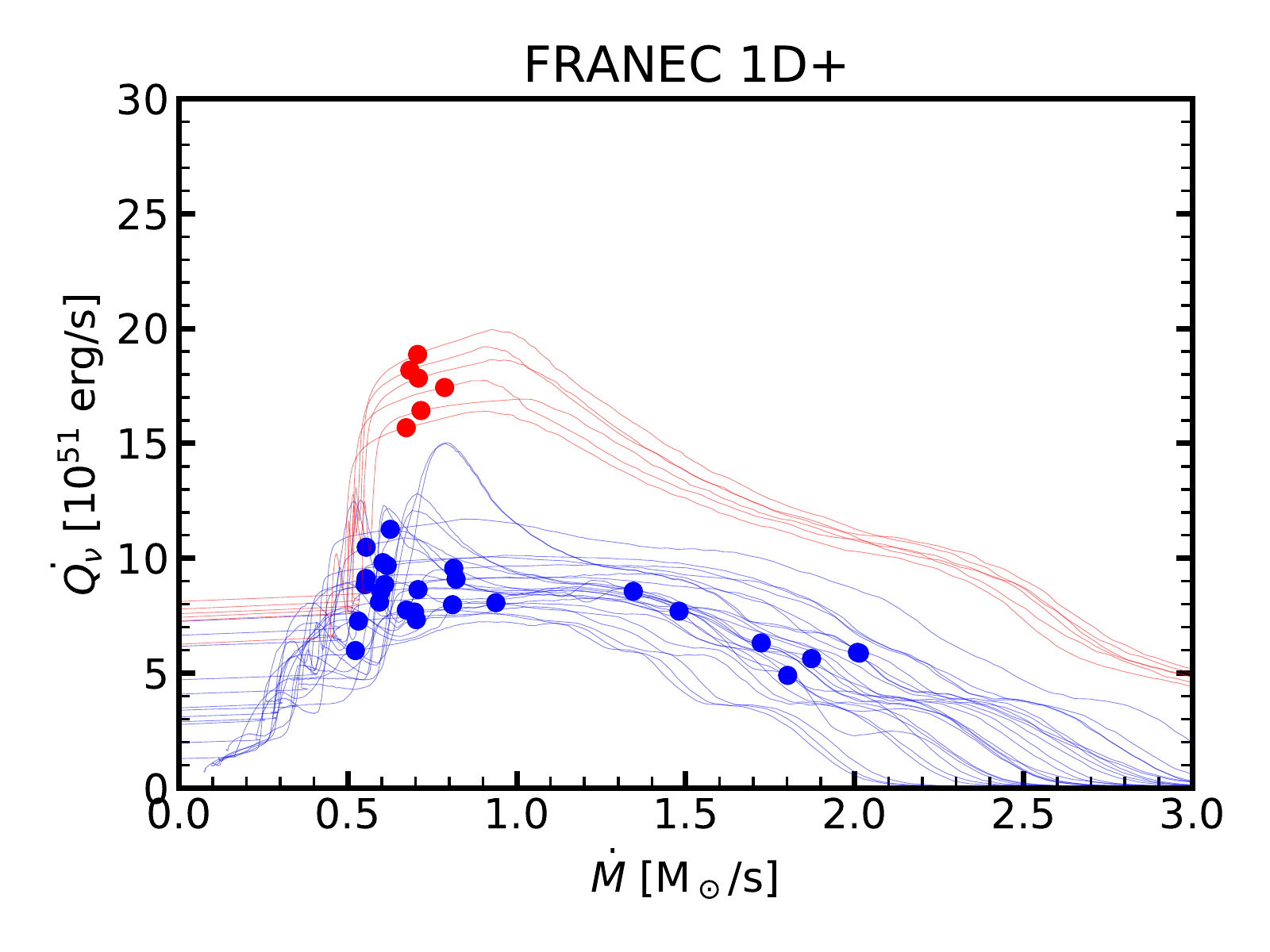}{0.5\textwidth}{(d)}}
    \caption{Same as Figure \ref{fig:maccr_heat_KEPLER} but for FRANEC progenitors.}
    \label{fig:maccr_heat_FRANEC}
\end{figure*}

The tracks in Figure \ref{fig:maccr_heat_KEPLER} show that the largest difference between 1D and 1D+ is for progenitors with $\xi_{2.5} > 0.3$.  This indicates how, for these progenitors, the contribution of $\nu$-driven convection is particularly important. Instead, in the $\dot{M}-L_\nu$ plane there would be no significant difference between 1D and 1D+. The same holds for the FRANEC progenitors, shown for completeness in Figure \ref{fig:maccr_heat_FRANEC}. The difference, as can be seen in the right panel of Figure \ref{fig:comp_vs_ZAMS_mass},  is that for FRANEC progenitors $\xi_{2.5} > 0.3$ corresponds to progenitor masses with $M > 25$ M$_\odot$. 

To summarize, it appears that in progenitors with large mass accretion rates, which can be achieved for large compactnesses $\xi_{2.5} > 0.3$, $\nu$-driven convection is very efficient. This is the reason why in previous studies, like the one by \cite{Ertl2016_explodability}, these progenitors did not explode. Their models did not account for this very important mechanism.

This analysis shows that different methods of triggering the explosion are more dependent on certain properties of the progenitor than others. In this case, the method of \cite{Ugliano2012} appears to strongly depend on $\mu_4$ and $M_4$, whereas our method strongly depends on $\delta\rho_*/\rho_*$. This might seem like a moot point, but it needs to be stressed that just because one method can produce an explosion using a physically motivated model, it does not mean that Nature operates in the same way. 

There are still many uncertainties in stellar evolution prescriptions.  These arise in part from using different codes \citep{Chieffi2020_presupernova_models} as well as different reaction rates \citep{Chieffi2021_C12_compactness,Sukhbold2020_missing_red_supergiant}. In addition to that, there are many uncertainties in the collapse and explosion phase affecting the nuclear matter equation of state, neutrino physics, and neutrino transport algorithms. 

In this work we only considered radial perturbations in density, that remain roughly constant during the accretion onto the shock. However, previous studies \citep{Kovalenko1998_instability_sph_accretion,Lai2000_growth_pert_during_accretion,Takahashi2014_non_sph_perturbations_preshock,Nagakura2019_semi_an_prog_asym} have shown that non-radial perturbations (i.e. $l > 0$) can significantly grow during accretion, with larger $l$ yielding larger amplifications. From this one can conclude that progenitor asphericities likely play a very important role in determining the outcome of the explosion, as shown by several 3D simulations \citep{Muller2016_Oxburning,OConnor2018_3Dprogenitors,Fields2021_3D_burning_precollapse}. 

Finally, in this study, we do not definitively claim that we can provide the true explodability of massive stars. By looking at the differences between the upper and bottom panels of Figure \ref{fig:calibration_explodability}, it is clear that large uncertainties in the stellar models can significantly affect the explodability of supernovae. Our goal, however, has been to show that $\nu$-driven convection plays an important role in reviving the shock together with the accretion of large density discontinuities.

\section{Conclusions}
\label{sec:conclusions}
In this paper, we have studied the pre-collapse properties that can best predict the explodability of core-collapse supernovae. To do that, we used STIR, a new model that simulates neutrino-heated turbulence in 1D CCSNe simulations. We have performed $\sim 1300$ core-collapse simulations of two progenitor sets from \cite{Sukhbold2016_explodability} and \cite{Chieffi2020_presupernova_models}. 

One of the main findings of this paper is that the outcome of the explosion is particularly sensitive to whether there is a density jump near the Si/Si-O interface. The accretion of the jump causes a decrease in the ram pressure, and therefore a surge of the shock, which causes the shock to temporarily expand. If the accretion occurs too early after bounce, when neutrinos are not yet generating significant heating, the surge has to be very large for the shock to break out. If this accretion occurs too late (i.e. $\gtrsim 400$ ms after bounce) then the shock will have already receded and the fate will always be to fall back onto the PNS and form a black hole.

First, we calibrated STIR using the results of \cite{Boccioli2022_EOS_effect}, as well as general considerations based on what values of $\alpha_{\rm MLT}$ yield explosion fractions compatible with observations. Then, we presented two criteria that predict whether a given progenitor explodes. Criterion (a) is based on dynamical quantities from the core-collapse simulations: (i) if $t_{\rm accr} > 0.4$ s, the star will not explode; (ii) if $t_{\rm accr} < 0.4$ s, the star will explode if $\delta R_{\rm surge}/R_{\rm accr} > 0.04$. Then, we showed that there is a correlation between $\delta R_{\rm surge}/R_{\rm accr}$ and $\delta \rho_*^2/\rho_*^2$, where $\rho_*$ is the density at which the density jump occurs, and $\delta \rho_*$ is the size of the jump. 

Therefore, criterion (b), which does not need any input from simulations, can be formulated: (i) if $\widetilde{t}_{\rm accr} > $ 0.4 s, the star won't explode, where $\widetilde{t}_{\rm accr}$ is defined in Equation \eqref{eq:t_accr_profile}; (ii) if $\widetilde{t}_{\rm accr} < 0.4$ s, the star will explode if $\delta \rho_*^2/\rho_*^2 > 0.08$. Criterion (a) yields $\sim$ 9.5 \% misclassifications, and criterion (b) yields $\sim$ 9 \% misclassifications. Criterion (b) can be used without the need of performing any core-collapse simulations, which makes it a powerful and robust tool to predict whether a given star will explode, based on its density profile.

A similar criterion, developed using 2D and 3D simulations, was recently published by \cite{Wang2022_prog_study_ram_pressure}. Our results agree extremely well, which is an indicator that criterion (b) is very robust since it can be obtained independently using multi-dimensional simulations as well as our 1D+ simulations. Moreover, the agreement between our 1D+ model and the results from \cite{Wang2022_prog_study_ram_pressure} confirms that the main ingredient missing from 1D simulations is $\nu$-driven turbulence, which is however included in our 1D+ simulations via STIR. 

Finally, we compared our results to \cite{Ertl2016_explodability}, who performed a similar study but used a different model to trigger the explosion. We analyzed the differences between the two methods and how they can affect the simulations, ultimately resulting in different explodabilities versus progenitor mass. The cause of these discrepancies can be traced to the fact that the models from \cite{Ertl2016_explodability} do not take the effects of $\nu$-driven turbulence into account, which are a very important ingredient, as shown in this paper.

A word of caution is in order, underlying the fact that large uncertainties still exist both in 1D and 3D models: both the EOS and neutrino opacities necessitate more accurate models, neutrino transport algorithms are still imperfect, the effects of GR are not yet fully understood, etc... This work points out another significant uncertainty that should not be underestimated: stellar evolution models. We showed that two different stellar evolution codes, KEPLER and FRANEC, yield different explosion patterns (see Figure \ref{fig:calibration_explodability}). Therefore, it is hard to determine whether a star of a given mass explodes or not since the uncertainties affecting stellar evolution propagate all the way through core-collapse.

Nonetheless, the agnostic nature of our criterion, based only on the density profile of the progenitor, proves that the most relevant feature of the progenitor star is the presence of a density jump near the Si/Si-O interface. Indeed, our criterion performs equally well on both FRANEC and KEPLER models. This shows that, although differences in stellar evolution prescriptions cause different explodabilities, the quantity to which the explosion is most sensitive remains the density jump near the Si/Si-O interface.

This criterion shows very good agreement with simulation results, with a success rate above $90\%$. Its simplicity and straightforward physical interpretation make it a robust tool that can be used to quickly predict the explodability of massive stars.

\acknowledgments
Work at the University of Notre Dame supported by DOE nuclear theory grant DE-FG02-95-ER40934.
ML and AC acknowledge the PRIN INAF 2019 ``From massive stars to supernovae and supernova remnants: driving mass, energy
and cosmic rays in our Galaxy" (P.I. S. Orlando, F.O. 1.05.01.85.02) and the Premiale Figaro 2015 (P.I. G. Ghirlanda, F.O. 1.05.06.13).
ML and AC thank both the Dipartimento di Fisica di Perugia and the Perugia INFN Unit for continuous and generous financial support.  

\bibliographystyle{aasjournal}
\bibliography{References_Stellar_Models,References_SN,References_CNO,References_Nucleosynthesis,References_EOS_neutrinos,References_Exp_Obs,References_Books}

\end{document}